\documentclass{aa}  
\usepackage[]{aalongtable}
\usepackage{scalefnt}
\usepackage{graphicx}
\usepackage{txfonts}
\begin{document}




\newbox\grsign \setbox\grsign=\hbox{$>$} \newdimen\grdimen \grdimen=\ht\grsign
\newbox\simlessbox \newbox\simgreatbox
\setbox\simgreatbox=\hbox{\raise.5ex\hbox{$>$}\llap
     {\lower.5ex\hbox{$\sim$}}}\ht1=\grdimen\dp1=0pt
\setbox\simlessbox=\hbox{\raise.5ex\hbox{$<$}\llap
     {\lower.5ex\hbox{$\sim$}}}\ht2=\grdimen\dp2=0pt
\def\simgreat{\mathrel{\copy\simgreatbox}}
\def\simless{\mathrel{\copy\simlessbox}}
\newbox\simppropto
\setbox\simppropto=\hbox{\raise.5ex\hbox{$\sim$}\llap
     {\lower.5ex\hbox{$\propto$}}}\ht2=\grdimen\dp2=0pt
\def\simpropto{\mathrel{\copy\simppropto}}

\title{ Oxygen and zinc abundances in 417 Galactic bulge red giants
\thanks{Observations collected at the European  Southern  Observatory,
  Paranal,  Chile  (ESO programmes  71.B-0617A, 73.B0074A);
Table B.1 is only available in electronic form at the CDS
via anonymous ftp to cdsarc.u-strasbg.fr (130.79.128.5) or
via http://cdsweb.u-strasbg.fr/cgi-bin/qcat?J/A+A/} }
\author{
C. R. da Silveira\inst{1}
\and
B. Barbuy\inst{1}
\and
A. C. S. Fria\c ca\inst{1}
\and
V. Hill\inst{2}
\and
M. Zoccali\inst{3,4}
\and
M. Rafelski\inst{5}
\and
O. A. Gonzalez\inst{6}
\and
D. Minniti\inst{4,7}
\and
A. Renzini\inst{8}
\and
S. Ortolani\inst{9}
}
\offprints{B. Barbuy}
\institute{
Universidade de S\~ao Paulo, IAG, Rua do Mat\~ao 1226,
Cidade Universit\'aria, S\~ao Paulo 05508-900, Brazil
\and
Universit\'e de Sophia-Antipolis,
 Observatoire de la C\^ote d'Azur, CNRS UMR 6202, BP4229, 
06304 Nice Cedex 4, France
\and
Pontificia Universidad Catolica de Chile, Instituto de Astrofisica,
Casilla 306, Santiago 22, Chile
\and
Millenium Institute of Astrophysics, Av. Vicu\~na Mackenna 4860,
Macul, Santiago, Chile
\and
Space Telescope Science Institute, 3700 San Martin Drive,
Baltimore, MD 21218, USA 
\and
Institute for Astronomy, University of Edinburgh, Royal Observatory,
Blackford Hill, Edinburgh, EH9 3HJ UK
\and
Departamento de Ciencias Fisicas, Universidad Andres Bello, 
Republica 220, Santiago, Chile
\and
Osservatorio Astronomico di Padova, Vicolo
 dell'Osservatorio 5, I-35122 Padova, Italy
\and
Universit\`a di Padova, Dipartimento di Astronomia, Vicolo
 dell'Osservatorio 2, I-35122 Padova, Italy
}

   \date{}

 
  \abstract
   {Oxygen and zinc in the Galactic bulge are key elements for 
the understanding of the bulge chemical evolution.
Oxygen-to-iron abundance ratios provide a most robust indicator of the 
 star formation rate and chemical evolution of the bulge. Zinc
is enhanced in metal-poor stars, behaving as an $\alpha$-element, and
its production may require nucleosynthesis in hypernovae.
 Most of the neutral gas at high redshift is in damped Lyman-alpha systems
 (DLAs), where Zn is also observed to behave as an $\alpha$-element.}
   {The aim of this work is the derivation of the $\alpha$-element oxygen, 
together with nitrogen, and the iron-peak element zinc 
abundances in 417 bulge giants, from moderate resolution (R$\sim$22,000)
 FLAMES-GIRAFFE spectra.  For stars in common with a set of UVES spectra
  with higher resolution (R$\sim$45,000), the data are intercompared.
 The
 results are compared with literature data and chemodynamical models.}
   {We studied the spectra obtained for a large sample of red giant stars,
chosen to be one magnitude above the horizontal branch,
 using  FLAMES-GIRAFFE on the 
Very Large Telescope. We computed the O abundances using the 
forbidden [OI] 6300.3 {\rm \AA}
and Zn abundances using the \ion{Zn}{I} 6362.34 {\rm \AA} lines. 
Stellar parameters for these stars were established in a previous work
from our group. }
   {We present oxygen abundances for 358  stars,
nitrogen abundances for 403 stars and zinc abundances were derived
for 333 stars. 
Having oxygen abundances for this large sample adds information in particular
at the moderate metallicities of -1.6$<$[Fe/H]$<$-0.8.
 Zn behaves as an $\alpha$-element, very similarly to
O, Si, and Ca.   It shows
 the same trend as a function of metallicity as the $\alpha$-elements, 
i.e., a turnover
around [Fe/H]$\sim$-0.6, and then decreasing with increasing metallicity.
The results are compared with chemodynamical evolution models 
of O and Zn enrichment for a classical bulge.
DLAs also show an enhanced zinc-to-iron ratio,
 suggesting they may be enriched by hypernovae. 
}
   {}


   \maketitle
%

\section{Introduction} 

Oxygen and zinc are key elements for the understanding of the
 star formation rate and
chemical enrichment of the Galactic bulge. Oxygen is the prime
and most robust probe for testing the timescale of 
bulge formation,  because it has no contribution from SNIa, and
 because the prescriptions from different authors
 (e.g. Woosley \& Weaver 1995, hereafter WW95; 
Kobayashi et al. 2006) produce the same behaviour.
Woosley et al. (2002) describe the nucleosynthesis production of
the different elements. 
In all cases oxygen is produced in hydrostatic phases of massive
star evolution.

Oxygen abundances in bulge field stars  have been derived in 
several studies, among which the most recent are
Alves-Brito et al. (2010), Bensby et al. (2013), Fria\c ca \& Barbuy (2017),
Johnson et al. (2014), J\"onsson et al. (2017), Mel\'endez et al. (2008),
Rich et al. (2012), Ryde et al. (2010),
Schultheis et al. (2017), and Siqueira-Mello et al. (2016).
A review on abundances in the Galactic bulge is given in McWilliam (2016).
A more general review on the MW bulge is presented in Barbuy et al. (2018).

Based on the observed [O/Fe] vs. [Fe/H] behaviour, as compared with their
chemodynamical models,
Cavichia et al. (2014), and  Fria\c ca \& Barbuy (2017, hereafter FB17), 
derived a specific star formation rate of
bulge formation and chemical enrichment of $\nu_{\rm SF}$ $\approx$ 0.5 Gyr$^{-1}$
or a timescale of bulge formation of 2 Gyr.
 The specific star formation is defined as
 $\nu_{\rm SF}$ = 1/M(M$_{\odot}$) dM(M$_{\odot}$)/dt,
which is the ratio of the SFR
 over the gas mass in M$_{\odot}$ available  for star formation. 
 A best value can be estimated from
the reproduction of the observed turnover in [O/Fe] vs. [Fe/H]
  by the models, that occur when SNIa start to give a contribution in Fe.

Zinc is key to probe the 
contribution of hypernovae at the lower metallicities during
the bulge chemical enrichment process. 
 The high [Zn/Fe] ratios in bulge metal-poor stars can at present only
 be explained
by enrichment from hypernovae (Kobayashi et al. 2006; Nomoto et al. 2013)
as discussed in Barbuy et al. (2015).
Zinc enhancements in metal-poor stars were derived in the literature also
 in halo stars by Cayrel et al. (2004),  Nissen \& Schuster (2011),
 in the thick disk by Bensby et al. (2014),
Reddy et al. (2006), Mishenina et al. (2011), 
and in metal-poor
 bulge stars by Bensby et al. (2013, 2017), and Barbuy et al. (2015).

 Zinc is also useful for comparisons
with data from  Damped Lyman-alpha systems (DLAs).
 DLAs are neutral hydrogen gas systems observed
 in absorption to background quasars, with minimum hydrogen column densities
 of 2$\times$10$^{20}$ cm$^{-2}$. DLAs dominate the neutral gas content at
 high redshift, and the metallicity in DLAs is observed to decrease with 
increasing redshift (Pettini 1999, Rafelski 2012, 2014), similar to the 
decrease of metallicity with age of stars in our Galaxy. Moreover, due to the 
neutrality of the gas, the metallicity of the gas can be measured quite 
precisely without ionization corrections, making them a premier site to
 measure abundances at high redshift. Additionally, DLAs have been found to
 be $\alpha$-enhanced and show enhanced [Zn/Fe] ratios (Rafelski et al. 2012).

We have previously studied the oxygen and zinc 
abundances in the Galactic bulge based on
high resolution FLAMES-UVES spectra of 56 bulge giants 
(Zoccali et al. 2006; Lecureur et al. 2007; 
 Barbuy et al. 2015;  FB17).
In the present work we derive O and Zn abundances for 417 red giants observed
with FLAMES-GIRAFFE, within the same observational programmes as the FLAMES-UVES
 data, at the Very Large Telescope.
The stars were observed in two fields, selected among the four fields
observed by Zoccali et al. (2008):
Baade's Window (BW) (l=1.14$^{\circ}$, b=-4.2$^{\circ}$), 
and a field at $\rm b=-6^{\circ}$ 
(l=0.2$^{\circ}$, b=$-6^{\circ}$).
These stars had already been analysed by
Zoccali et al. (2008), and  Gonzalez et al. (2011)
 derived abundances of the $\alpha$-elements
Mg, Si, Ca, and Ti for the sample.

The sample covers a range in metallicity [Fe/H] that allows us to investigate 
the bulge chemical evolution history in connection to other Galactic 
components. It includes 65 stars with [Fe/H]$\leq$-0.5, and
14 with [Fe/H]$\leq$-1.0, thus covering the required range, 
 to help  impose fundamental constraints 
on chemical enrichment models from oxygen and zinc abundances.
 This is so because the bulk of the bulge
stars cover the metallicity range of $\sim$-1.3$<$[Fe/H]$<$$\sim$+0.5
(Hill et al. 2011, Ness et al. 2013, 
Rojas-Arriagada et al. 2017, Zoccali et al. 2017),
so that stars with metallicities in the range of $-$1.3$<$[Fe/H]$<$$-$0.8
are important to understand the metal-poor end of the bulge
chemical enrichment.
 A metal-poor end at these relatively high metallicities
can be explained by the fast chemical enrichment
that takes place in the bulge, rapidly reaching the metallicity
of [Fe/H]$\sim$-1.0 (e.g. Cescutti et al. 2008, Wise et al. 2012,
Cescutti et al. 2018). In other words, the equivalent
of [Fe/H]$\sim$-3.0 in the halo, is [Fe/H]$\sim$-1.0 in the bulge.
It can also help to 
better constrain the interfaces between the old bulge with the
inner halo, and the thick disk. 
 Hawkins et al. (2015) suggested that moderately 
metal-poor stars in the bulge could define the interface of
Galactic disk and inner halo, by studying stars within
$-$1.20$<$[Fe/H]$<$$-$0.55. The connection between thin and thick disks has also been studied by Mikolaitis et al. (2014).

 In the present paper we have adopted the chemodynamical evolution
models for an old classical bulge
 described in Barbuy et al. (2015) and FB17, with some
modifications.
In Sect. 2 the observations are summarized. 
In Sect. 3 the basic stellar parameters are reported, and the
abundance derivation of O and Zn is described.
In Sect. 4 results and discussion are presented,
including comparison with literature and
 chemodynamical evolution models. In Sect. 5
O-poor and N-rich field stars are selected. 
A summary is given in Sect. 6.


\section{Observations}
The present  data were obtained using the FLAMES-GIRAFFE
instrument at the 8.2 m Kueyen of the Very Large Telescope,
at the European Southern Observatory (ESO), in Paranal, Chile,
 as described in 
Zoccali et al. (2008) (ESO Projects 071.B-067, 071.B-0014; PI: A. Renzini).
Targets are bulge K giants, with magnitudes $\sim$1.0 above the
red clump, originally in four fields.

In the present work, we have analysed stars from two fields, 
reported in Table \ref{fields},
 among the four fields studied in Zoccali et al. (2008). 
These two fields were observed in setups
 HR13 (612.0 - 640.5nm), HR14A (630.8 - 670.1nm)
 and HR15 (660.7 - 696.5nm), with
  resolving power respectively of R = 26,400, 18,000, and 21,350.
The other two fields, NGC 6553 field
at (l,b) =  (5\fdg25,−3\fdg02) and Blanco field
at (l,b) =   (0$^{\circ}$,−12$^{\circ}$),
observed with setups HR11, HR13, and HR15, were reanalysed
 by Johnson et al. (2014),  chosen by them because the HR11 setup 
 contains copper lines. They derived abundances of the light elements Na, Al,
$\alpha$-elements O, Mg, Si and Ca, and the Fe-peak
elements Cr, Fe, Co, Ni, and Cu, for 156 red giants in those fields.

\begin{table}[ht!]
\label{fields}
\caption{Fields observed: coordinates, distance to Galactic centre,
reddening  as adopted in Zoccali et al. (2008), 
number of stars observed, and typical signal-to-noise ratios.} 
\scalefont{1.05}
\[
\begin{array}{|lc@{}c@{}ccccc|}
\hline\hline
\noalign{\smallskip}
\hbox{Field} & \hbox{l} & \hbox{b}
& \hbox{R$_{\rm GC}$} & \hbox{E(B-V)} & 
\hbox{N$_{\rm stars}$} & \hbox{S/N/pixel} & \\
\hbox{} & \hbox{($^{\circ}$)} & \hbox{($^{\circ}$)}
& \hbox{(pc)} & \hbox{} & 
\hbox{} & \hbox{@620nm} & \\
\noalign{\smallskip}
\hline
\noalign{\smallskip}
\hbox{Baade's window}  &  1\fdg14  &  \phantom{-}-4\fdg18  & 604 & 0.55 & 204 &40-60& \\
\hbox{b=-6$^{\circ}$}  &   0\fdg21   &  \phantom{-}-6\fdg02  & 850 & 0.48 & 213&60-90& \\
\noalign{\smallskip}
\hline
\end{array}
\]
\end{table}

Given that the same stars observed  with FLAMES-UVES were also observed
with FLAMES-GIRAFFE, the oxygen and zinc were derived also from the
GIRAFFE spectra. The fits
to both UVES and GIRAFFE spectra are shown in Appendix A for the stars in common between the two.
 
The sample consists of red giant branch (RGB) stars, chosen to be
$\sim$1.0 magnitude above the horizontal branch, consequently 
 this sample does not include red clump (RC) stars, as is the case of more
recent surveys (e.g. Ness et al. 2013; Rojas-Arriagada et al. 2017; Zoccali
et al. 2017). Such a selection was intended to
 exclude brighter RGB stars in order to avoid
spectra with strong TiO lines. The V, I and astrometric positions are from
the OGLE catalogue (Udalski et al. 2002), a pre-FLAMES catalogue
 (Momany et al. 2001), and 2MASS (Carpenter 2001), as described in
Zoccali et al. (2008).
 The stars' names follow the observational strategy:
the targets were divided into two samples, bright and faint, in order to
optimize exposure time. When a sample was being observed with GIRAFFE,
the other one was observed with UVES, and then the two were swapped.
 The total exposure time
varied from about 1 h to almost 5 h, depending on the setup and
on the star luminosity, in order to ensure that the final S/N per pixel,
 of each co-added spectrum to reach $\sim$60 
(see mean S/N per field in Table \ref{fields}).
 Therefore the identifications
are Baade's window bright (BWb) and faint (BWf), and the same
for the -6 degree field with bright stars identified by B6b, and
faint ones by B6f. These identifications for the UVES stars are inverted
 for the GIRAFFE identifications,  that is, a BWb or B6b star in UVES will
be a BWf or B6f in GIRAFFE, with numbers at random, corresponding
to a random allocation of fibres for the observations with GIRAFFE.


\section{Abundance analysis}

Elemental abundances were obtained through line-by-line spectrum synthesis
calculations, carried out 
using the code described in Barbuy et al. (2003) and Coelho et al. (2005).
The main molecular lines present in the region, namely the 
CN  B$^2$$\Sigma$-X$^2$$\Sigma$ blue system,
 CN  A$^2$$\Pi$-X$^2$$\Sigma$ red system,
 C$_2$  Swan A$^3$$\Pi$-X$^3$$\Pi$, MgH A$^3$$\Pi$-X$^3$$\Sigma^{+}$,
 and TiO A$^3$$\Phi$-X$^3$$\Delta$ $\gamma$ and
B$^3$$\Pi$-X$^3$$\Delta$ $\gamma$' systems were taken into account.
The atmospheric models were obtained by interpolation in the grid
of spherical and mildly CN-cycled
 ([C/Fe] = −0.13, [N/Fe] = +0.31) MARCS models by Gustafsson et al. (2008).
 These models consider 
[$\alpha$/Fe] = +0.20. 
  These models were chosen as this value is compatible with
the C, N values in normal red giants, and have suitable $\alpha$-element
 enhancements.

We adopted the stellar parameters established  by our group, given in
 Zoccali et al. (2006, 2008), and reported in Table \ref{results}.
 A brief description of the methods follows:

\begin{itemize}

\item Photometric colours [{\rm \tt V,I}] were used together with
colour-temperature calibrations by Ram\'{\i}rez \& Mel\'endez (2005).
Another useful indicator was also used: the intensity of TiO bands.
Given that RGB stars were chosen, intentionally not very bright, in
order to avoid too strong TiO bands, it was possible to 
define a TiO band index, measuring its strength at 6190-6250 {\rm \AA},
for stars with T$_{\rm eff}$$<$ 4500 K (see Zoccali et al. 2008 for further
details).
Effective temperatures were then checked by imposing excitation
equilibrium for FeI and FeII lines of different excitation potential,
using about 60 FeI lines, selected to be suitable for metallicities
down to [Fe/H]$\sim$-0.8, and another line list for more metal-poor stars.

Since the final temperatures are spectroscopic, 
the reddening E(B-V) and photometric temperatures were used only
as initial guesses. The values of reddening reported in Table 1
are from Zoccali et al. (2006, 2008), and are compatible
with the minimum values given in Schlafly \& Finkbeiner (2011) in
fields of 2$^{\circ}$\footnote{http://irsa.ipac.caltech.edu/applications/DUST/}.
\item Photometric gravities of the sample stars were obtained 
adopting a classical relation,   where the bolometric corrections were
obtained using relations by Alonso et al. (1999). 

\item Microturbulent velocities v$_{\rm t}$ were determined by imposing
a constant [Fe/H] derived from FeI lines of different
expected equivalent widths.

\item Finally, the metallicities for the sample stars 
were derived using a set of
equivalent widths of \ion{Fe}{I} lines. 

\end{itemize}

These stellar parameters were also adopted by Gonzalez et al. (2011),
for the derivation of $\alpha$-element abundances.
For stars in common with the UVES data, the stellar parameters
derived by Zoccali et al. (2006) from FLAMES-UVES data were used.

In Zoccali et al. (2006) and Lecureur et al. (2007), the oxygen abundance
for the 56 giants observed with FLAMES-UVES
 were derived. In Barbuy et al. (2015) these
values were revisited, with the unique aim of obtaining reliable CN strengths.
In FB17 the oxygen abundances in stars of this sample,
observed with both UVES and GIRAFFE spectrographs, were further
 revised by taking
into account in more detail the abundances of carbon based on the
 C$_2$(0,1) bandhead at 5635.2 {\rm \AA} and
the \ion{C}{I} 5380.3 {\rm \AA} line. 
 These derivations replace the previous
values by Zoccali et al. (2006), and Lecureur et al. (2007).
 A mean [C/Fe]=$-$0.07$\pm$0.09 was found
for the UVES sample.
Recently, J\"onsson et al. (2017)  and Schultheis et al. (2017)
reanalysed a fraction of the FLAMES-UVES sample (see Sect. 
\ref{literatureox}).

\subsection{Zinc}
In Barbuy et al. (2015), we derived zinc abundances for 56 red giants 
observed with
the FLAMES-UVES spectrograph. The \ion{Zn}{I} 4810.53 and 6362.34 {\rm \AA}
lines were used to derive the zinc abundances.
 The sample in the present work contains 23 stars 
observed with UVES. We revised the Zn abundances from the
 \ion{Zn}{I} 4810.53 {\rm \AA} line observed  with UVES for stars in common
with the present sample. The abundances from Barbuy et al. (2015) are reported
in Table \ref{giraffeuves}. In a few cases a corrected value is indicated
in bold face.

 In the present work, 
the FLAMES-GIRAFFE spectra contain
the  \ion{Zn}{I}  6362.34 {\rm \AA} line alone.
As mentioned above, in Appendix A the fits to this line with
both UVES and GIRAFFE spectra are shown, for the stars common to
the two samples.
 Literature and adopted oscillator strengths 
were reported, together with blending lines in Table 1 of Barbuy et al. (2015).
The effect of  a continuum lowering in the range
$\sim$6360.8 - 6363.1 {\rm \AA},
 due to the \ion{Ca}{I} 6361.940 autoionization line was taken into account.
The continuum in the range 6361-6362 {\rm \AA} was the prime
reference for fitting the Zn line, where the effects of the
\ion{Ca}{I} autoionization line put this region and the \ion{Zn}{I} line
at the same continuum level.
The FWHM of lines was fitted for each star for a region around the
 \ion{Zn}{I} line.

The \ion{Zn}{I} 6362.339 {\rm \AA} line is sometimes blended with CN lines,
as extensively discussed in Barbuy et al. (2015).
For this reason, it is necessary to have a suitable derivation of 
C, N, and O abundances. In the present work we have derived Zn abundances for 333 stars among the 417
sample ones, where the line was well defined.

\subsection{Carbon, nitrogen, and oxygen abundances}

The derivation of C, N, and O abundances proceeded as described below. 

Carbon: since the present spectra have neither the  
Swan C$_2$ (0,1) A$^3$$\Pi$-X$^3$$\Pi$ bandhead at
5635 {\rm \AA}, nor the \ion{C}{I} 5380.3 {\rm \AA} line,
 and in the absence of a reliable C abundance indicator, we adopted
a value of [C/Fe]=$-$0.2 for all stars,
 a deficiency expected in red giants
 (e.g.  Smiljanic et al. 2009),
 compatible with the mean [C/Fe]=$-$0.07 found for the UVES sample,
see FB17, their Table A.1.
For stars that are also observed with UVES, as well as with 
Zoccali et al. (2006), Barbuy et al. (2015), and FB17,
the UVES results are preferred.
 The effect of C abundance in the O
abundance is illustrated in Barbuy (1988, their Fig. 2).
The N abundance, as derived from a CN bandhead depends on the C 
abundance adopted. Despite an uncertainty on the N abundance due
to this assumption, we 
 remind the reader that the main aim here is to be able to reproduce
the CN line intensities, given the blend with CN lines on the
right wing of the  \ion{Zn}{I} 6362.339 {\rm \AA} line.

Nitrogen: we used  the red CN (5,1) 
A$^2$$\Pi$-X$^2$$\Sigma$ bandhead
at 6332.18 {\rm \AA}, to derive N abundances, adopting the laboratory  
line list by Davis \& Phillips (1963).
Nitrogen abundances are important for the dissociative 
equilibrium between C,N, and O  (e.g. Tsuji 1973; Irwin 1988).
 In red giants N abundances are more informative on the CN-cycle
than on chemical evolution.  This is due to the
transformation of C into N due to  the CNO-cycle 
that takes place along the ascent of the giant branch.
Added to the expected mixing process, there is an
observed extra-mixing (see e.g. Smiljanic et al. 2009).
Therefore the enhanced N abundances observed are due to
stellar evolution processes, and do not reflect necessarily
the N abundance of the gas from which the star formed.
 The very few cases of 
very high nitrogen abundances, combined
with low oxygen abundances, are discussed in Sect. \ref{nrich}.

Oxygen: the forbidden oxygen [OI]6300.311 {\rm \AA} line was used
to derive O abundances, adopting log gf = -9.716, and taking
into account the blends with \ion{Ni}{I} lines at 6300.300 and 6300.350
{\rm \AA},  where we adopted Ni abundances varying in lockstep
with Fe, as expected (e.g. Bensby et al. 2014, 2017).
A solar abundance of A(O)=8.76 is adopted (Steffen et al. 2015).

 In conclusion, the abundances of N, O and Zn were derived
iteratively in this order. 
 The CN
line intensity that appears as an  asymmetry on the right wing of the
 \ion{Zn}{I} 6362 {\rm \AA} line, is also used to check the N and O abundances.

Table \ref{results} gives the atmospheric parameters adopted from Zoccali et al.
(2008), and the resulting N, O, and Zn abundances
for 417 stars in Baade's window and the -6 degree fields.
In Table \ref{giraffeuves} are presented the abundances derived
for N, O, and Zn for the 23 sample stars having FLAMES-UVES spectra. The
Zn abundances from the ZnI 4810 {\rm \AA} line were revised, and
slightly modified in a few cases (indicated in bold face).
For deriving the
present N, O, and Zn abundances for these stars,
we adopted the parameters from the UVES analysis (Zoccali et al. 2006;
Lecureur et al. 2007). As explained above (Sect. 3),
the C, N, and O abundances 
reported first in Zoccali et al. (2006) and Lecureur et al. (2007),
were partially revised in Barbuy et al. (2015), and the revision was
further completed by FB17, and these latter are the values adopted here. 

\begin{figure}
\centering
\includegraphics[width=\columnwidth]{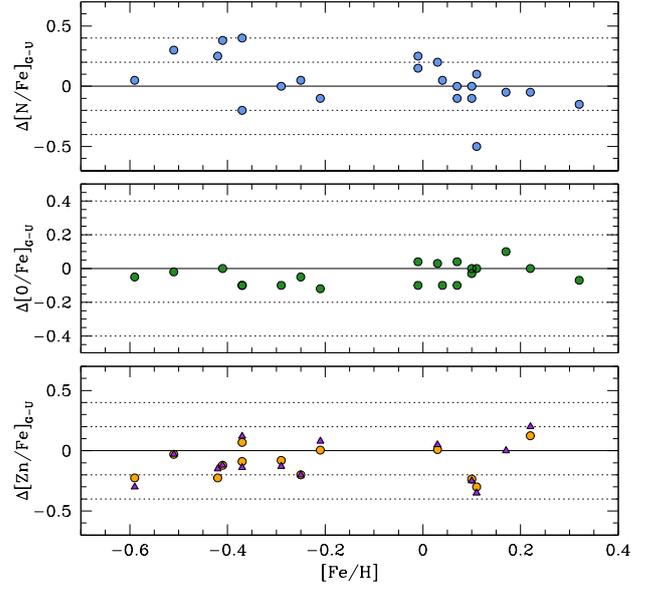}
\caption{Comparison between GIRAFFE and UVES abundances for
O, N, and Zn for the stars common to the two sets of spectra
 (Table \ref{both}).
For Zn: orange full circles consider the mean of two lines
in the UVES results, and violet full triangles compare results
for the same line. }
\label{giraffe-uves} 
\end{figure}

 In Table \ref{both} we report the stellar parameters 
and N, O, and Zn abundances
for the 23 stars from both UVES and GIRAFFE data. 
In Fig. \ref{giraffe-uves} we compare the abundances of O, N, and Zn
derived from the GIRAFFE data with those derived from the UVES spectra.
The oxygen abundances are in very good agreement. Nitrogen abundances
appear somewhat higher in GIRAFFE spectra with respect to those in UVES.
For N we could not refit C, since we have no atomic or molecular line
for this element, and this may be the source of the discrepancy.
Zinc tends to be lower in GIRAFFE  spectra than in the UVES ones;
 in Fig.  \ref{giraffe-uves} the difference is larger for
the UVES values given for the mean of abundances derived from the two lines 
 \ion{Zn}{I} 4810.5 and 6362.3 {\rm \AA}, and less discrepant when
comparing results for the same line as in the GIRAFFE spectra.

\subsection{Errors}

 For stars in common with UVES, we adopted the same
uncertainties given in 
Barbuy et al. (2013), amounting to
T$_{\rm eff}$  $\pm$ 150 K for effective temperature, 
log g $\pm$ 0.20 for surface gravity,  
[Fe/H] $\pm$ 0.10 in metallicity, 
 and v$_{\rm t}$ $\pm$ 0.10 kms$^{-1}$ for microturbulent velocity.
For the stars that have only GIRAFFE spectra we adopted higher uncertainties,
 due to having a lower resolution in the measurements of 
\ion{Fe}{I} and \ion{Fe}{II} lines,
of $\pm$ 200 K for T$_{\rm eff}$,
 $\pm$ 0.40 for log g,  
$\pm$ 0.10 in [Fe/H]
 and $\pm$ 0.30 kms$^{-1}$ for microturbulent velocity.

 The errors in [O/Fe] and [Zn/Fe] are computed by using model atmospheres
with parameters changed by these uncertainties, applied to the 
representative stars: the cooler star BW-b6, and the hotter star
B6-b3. Both of these were also analysed by J\"onsson et al. (2017),
 as shown in Table \ref{jonsson}.

 These uncertainties are given in Table \ref{errors2}. 
Since the stellar parameters are covariant, the sum of these errors
is an upper limit.
On the other hand, a continuum location uncertainty introduces a
further uncertainty in [O/Fe]$\sim$$\pm$0.05 and [Zn/Fe]$\sim$$\pm$0.05.

\begin{table*}
\caption{Sample of stars observed with both FLAMES-UVES and FLAMES-GIRAFFE.
Metallicities [Fe/H] are from Zoccali et al. (2006), 
 N, O abundances are from Fria\c ca \& Barbuy (2017).
 [Zn/Fe] from Barbuy et al. (2015) for \ion{Zn}{I} 4810.54 {\rm \AA}
and if revised they
are indicated in bold face; [Zn/Fe] for the \ion{Zn}{I} 6362.3 {\rm \AA} 
line in both UVES and GIRAFFE spectra are shown in Appendix A.}             
\label{giraffeuves} 
\scalefont{1.0}     
\centering          
\begin{tabular}{lrrrrrrrrrrrrrrrrrrrrr}     
\noalign{\vskip 0.1cm}    
\noalign{\vskip 0.1cm} 
\hline\hline    
\noalign{\smallskip}
GIRAFFE & UVES  & [Fe/H]& [N/Fe]& [O/Fe]& [Zn/Fe]   &[Zn/Fe] & $<$[Zn/Fe]$>$\\ 
\noalign{\vskip 0.1cm} 
\hline    
\noalign{\smallskip}
        &   & & & & \hbox{(ZnI 4810 \AA)}    &\hbox{(ZnI 6362 \AA)}&   & \\ 
\noalign{\vskip 0.1cm} 
\hline
\noalign{\smallskip}
bwb007  & BW-f1 & +0.32 & +0.45 & -0.18 &   -0.35      & -0.30 & $-$0.30 &    \\ 
bwb040  & BW-f5 & -0.59 & +0.40 & +0.25 &   {\bf +0.30}& +0.00 &   +0.15 \\ 
bwb061  & BW-f7 & +0.11 & +0.70 & -0.25 &   -0.20      & +0.00 &   $-$0.10 \\ 
bwb087  & BW-f4 & -1.21 & +0.70 & +0.30 &   +0.30      & +0.00 &   +0.15 \\ 
bwb096  & BW-f6 & -0.21 & +0.40 & +0.20 &   +0.15      & +0.00 &   +0.08 \\ 
bwf026  & BW-b5 & +0.17 & +0.05 & -0.10 &  -0.30       & -0.30 &   $-$0.30 \\ 
bwf067  & BW-b2 & +0.22 & +0.20 & -0.10 &   {\bf -0.30}& +0.00 &   $-$0.15 \\ 
bwf093  & BW-b4 & +0.07 & +0.00 & -0.10 &   {\bf -0.60}& -0.30 &   $-$0.45 \\ 
bwf102  & BW-b6 & -0.25 & +0.65 & +0.15 &   +0.00      & +0.00 &   +0.00 \\ 
bwf119  & BW-b7 & +0.10 & +0.10 & -0.20 &   {\bf -0.30}& -0.30 &   $-$0.30 \\ 
b6b044  & B6-f5 & -0.37 & +0.00 & +0.10 &   +0.10      & +0.00 &   +0.05 \\ 
b6b060  & B6-f7 & -0.42 & +0.30 & ---   &   -0.15      & +0.00 &   $-$0.08 \\ 
b6b095  & B6-f2 & -0.51 & +0.20 & +0.20 &   +0.05      & +0.00 &   +0.03 \\ 
b6b122  & B6-f1 & -0.01 & +0.20 & +0.03 &   -0.30      & -0.60 &   $-$0.45 \\ 
b6b132  & B6-f8 & +0.04 & +0.30 & -0.20 &   -0.60      & -0.30 &   $-$0.45 \\ 
b6b134  & B6-f3 & -0.29 & +0.30 & +0.15 &   +0.10      & +0.00 &   +0.05 \\ 
b6f010  & B6-b1 & +0.07 & +0.50 & +0.00 &   -0.20      & -0.30 &   $-$0.25 \\ 
b6f013  & B6-b8 & +0.03 & +0.10 & -0.03 &   {\bf -0.30}& +0.00 &   $-$0.15 \\ 
b6f016  & B6-b3 & +0.10 & +0.50 & -0.12 &   -0.27      & -0.60 &   $-$0.44 \\ 
b6f028  & B6-b5 & -0.37 & +0.30 & +0.15 &   {\bf -0.15}& +0.00 &   $-$0.08 \\ 
b6f062  & B6-b2 & -0.01 & +0.35 & +0.00 &   -0.15        & ---   &   $-$0.15 \\
b6f092  & B6-b4 & -0.41 & +0.15 & +0.30 &   +0.00      & +0.00 &   +0.00 \\ 
b6f095  & B6-b6 & +0.11 & +0.50 & -0.10 &  {\bf -0.30} & -0.30 &   $-$0.30 \\ 
\hline                   
\hline                  
\end{tabular}
\end{table*}  

\begin{table*}
\caption{Sample of 23 stars observed with both FLAMES-UVES, 
and FLAMES-GIRAFFE for comparison purposes.}             
\label{both} 
\scalefont{0.8}     
\centering          
\begin{tabular}{l@{}r@{}rrrrrrrrrrrrrrrr}     
\noalign{\vskip 0.1cm}
\noalign{\hrule\vskip 0.1cm}
\noalign{\vskip 0.1cm}    
 & \multicolumn{8}{c}{\hbox{\bf UVES}} 
& & \multicolumn{5}{c}{\hbox{\bf GIRAFFE}} &  \\
\noalign{\vskip 0.1cm} 
\hline\hline    
\noalign{\smallskip}
 OGLE & UVES  & GIRAFFE & T$_{\rm eff}$ & log~g & [Fe/H]& v$_{\rm t}$ & [N/Fe]& [O/Fe]&[Zn/Fe]& 
 T$_{\rm eff}$ & log~g & [Fe/H]& v$_{\rm t}$& [N/Fe]& [O/Fe]& [Zn/Fe] & \\

\hline\hline    
\noalign{\smallskip}
433669  & BW-f1 &       bwb007  &       4400    &       1.80    &       0.32    &       1.6     &       0.45    &       -0.18   &       -0.30   & 
        4300    &       1.67    &       0.32    &       1.5     &       0.30    &       -0.25   &       -0.40   & \\
240260  & BW-f5&        bwb040  &       4800    &       1.90    &       -0.59   &       1.3     &       0.40    &       0.25    &       0.00    & 
        5150    &       2.07    &       -0.59   &       1.4     &       0.45    &       0.20    &       0.30    & \\
357480  & BW-f7&        bwb061  &       4400    &       1.90    &       0.11    &       1.7     &       0.70    &       -0.25   &       -0.10   &         4800    &       2.06    &       0.11    &       1.4     &       0.20    &       -0.25   &       -0.15   & \\
537070  & BW-f4&        bwb087  &       4800    &       1.90    &       -1.21   &       1.7     &       0.70    &       0.30    &       0.15    &         5150    &       2.14    &       -1.21   &       1.1     &       0.54    &       0.30    &       ----    & \\
392918  & BW-f6&        bwb096  &       4100    &       1.70    &       -0.21   &       1.5     &       0.40    &       0.20    &       0.08    &         4600    &       1.97    &       -0.21   &       1.4     &       0.30    &       0.08    &       0.22    & \\
82760   & BW-b5&        bwf026  &       4000    &       1.60    &       0.17    &       1.2     &       0.05    &       -0.10   &       -0.30     &       4300    &       1.87    &       0.17    &       1.5     &       0.00    &       0.00    &       ----    & \\
214192  & BW-b2&        bwf067  &       4300    &       1.90    &       0.22    &       1.5     &       0.20    &       -0.10   &       -0.15   &         4450    &       1.96    &       0.22    &       1.3     &       0.15    &       -0.10   &       -0.20   & \\
545277  & BW-b4&        bwf093  &       4300    &       1.40    &       0.07    &       1.4     &       0.00    &       -0.10   &       -0.45   &         4100    &       1.84    &       0.07    &       1.2     &       -0.10   &       -0.20   &       ----    & \\
392931  & BW-b6&        bwf102  &       4200    &       1.70    &       -0.25   &       1.3     &       0.65    &       0.15    &       0.00    &         4450    &       1.89    &       -0.25   &       1.5     &       0.70    &       0.10    &       0.20    & \\
554694  & BW-b7&        bwf119  &       4200    &       1.40    &       0.10    &       1.2     &       0.10    &       -0.20   &       -0.30   &         4300    &       1.89    &       0.10    &       1.2     &       0.10    &       -0.20   &       -0.10   & \\
33058c2 & B6-f4&        b6b044  &       4500    &       1.80    &       -0.37   &       1.4     &       0.00    &       0.10    &       0.05    &         4550    &       1.84    &       -0.37   &       1.7     &       0.40    &       0.00    &       0.08    & \\
100047c6& B6-f7&        b6b060  &       4300    &       1.70    &       -0.42   &       1.6     &       0.30    &       0.25    &       -0.08   &         4350    &       1.72    &       -0.42   &       1.5     &       0.55    &       0.25    &       0.15    & \\
90337c7& B6-f2& b6b095  &       4700    &       1.70    &       -0.51   &       1.5     &       0.20    &       0.20    &       0.03    &         4850    &       2.02    &       -0.51   &       1.5     &       0.50    &       0.18    &       0.08& \\
23017c3& B6-f1& b6b122  &       4200    &       1.60    &       -0.01   &       1.5     &       0.20    &       0.03    &       -0.45   &         4250    &       1.65    &       -0.01   &       1.5     &       0.35    &       0.07    &       0.00    & \\
11653c3& B6-f8& b6b132  &       4900    &       1.80    &       0.04    &       1.6     &       0.30    &       -0.20   &       -0.45   &         4850    &       1.91    &       0.04    &       1.5     &       0.35    &       -0.30   &       -0.50   &\\
21259c2& B6-f3& b6b134  &       4800    &       1.90    &       -0.29   &       1.3     &       0.30    &       0.15    &       0.05    &         5000    &       2.02    &       -0.29   &       1.5     &       0.30    &       0.05    &       0.13    & \\
29280c3& B6-b1& b6f010  &       4400    &       1.80    &       0.07    &       1.6     &       0.50    &       0.00    &       -0.25   &         4350    &       1.80    &       0.07    &       1.5     &       0.50    &       0.04    &       ----    & \\
108051c7& B6-b8&        b6f013  &       4100    &       1.60    &       0.03    &       1.3     &       0.10    &       -0.03   &       -0.15   &         4250    &       1.79    &       0.03    &       1.6     &       0.30    &       0.00    &       -0.05   & \\
31220c2& B6-b3& b6f016  &       4700    &       2.00    &       0.10    &       1.6     &       0.50    &       -0.12   &       -0.44   &         4400    &       1.81    &       0.10    &       1.7     &       0.40    &       -0.15   &       -0.05   & \\
31090c2& B6-b5& b6f028  &       4600    &       1.90    &       -0.37   &       1.8     &       0.30    &       0.15    &       -0.08   &         4700    &       1.98    &       -0.37   &       1.5     &       0.10    &       0.05    &       0.14    & \\
83500c6& B6-b2& b6f062  &       4200    &       1.50    &       -0.01   &       1.4     &       0.35    &       0.00    &       -0.15   &         4400    &       2.00    &       -0.01   &       1.4     &       0.60    &       -0.10   &       ----    & \\
60208c7& B6-b4& b6f092  &       4400    &       1.90    &       -0.41   &       1.7     &       0.15    &       0.30    &       0.00    &         4400    &       1.83    &       -0.41   &       1.6     &       0.53    &       0.30    &       0.12    & \\
77743c7& B6-b6& b6f095  &       4600    &       1.90    &       0.11    &       1.8     &       0.50    &       -0.10     &       -0.30   &       4350    &       1.78    &       0.11    &       1.5     &       0.60    &       ----    &       -0.15   & \\
\hline                   
\hline                  
\end{tabular}
\end{table*}

\begin{table*}
\scalefont{0.9} 
\caption{Uncertainties on the derived [O/Fe] and
 [Zn/Fe] values for model changes of $\Delta$T$_{\rm eff}$ = 150, 200 K,
$\Delta$log g = +0.2, 0.4, 
 $\Delta$v$_{\rm t}$ = +0.1, 0.2, 0.3 km s$^{-1}$, for UVES and GIRAFFE data respectively,
  and corresponding total error, applied to the
stellar parameters T$_{\rm eff}$, log~g, [Fe/H], v$_{\rm t}$ of
stars BW-b6 (4200 K, 1.7, $-$0.25, 1.3 km.s$^{-1}$),
and B6-b3 (4700 K, 2.0, 0.10, 1.6 km.s$^{-1}$). 
The errors are given such
as the difference is the amount needed to recover the correct fit.
} 
\label{errors2}
\[
\begin{array}{l@{}c@{}c@{}c@{}c@{}c@{}c@{}cc}
\hline\hline
\noalign{\smallskip}
\hbox{Star} & \hbox{Element} &
 \hbox{$\Delta$T$_{\rm eff}$} & \hbox{$\Delta$$\log$ g}
& \hbox{$\Delta$v$_{t}$} 
& \hbox{($\sum$x$^{2}$)$^{1/2}$ }  & \hbox{continuum} & \hbox{($\sum$x$^{2}$)$^{1/2}$ } \\
& & \hbox{(+150 K)} & \hbox{(+0.2)} &  \hbox{(+0.1 kms$^{-1}$})& {\rm (parameters)} &
 &    {\rm (final)} &\\
\noalign{\smallskip}
\hline
\noalign{\smallskip}
\hbox{UVES} &\hbox{[C/Fe](CI)} &  +0.00  & +0.00 & +0.00 & +0.00  & \pm{0.02}  & {0.02}  & \\
\hbox{BW-b6}&\hbox{[C/Fe](CH)} &  +0.00  & +0.00 & +0.00 & +0.00  & \pm{0.05}  & {0.05} & \\
     &\hbox{[N/Fe]}     &  -0.08  & +0.02 & +0.00 & +0.08  & \pm{0.02}  & {0.08} & \\
     & \hbox{[O/Fe]}    &  +0.00  & +0.02 & +0.00 & +0.02  & \pm{0.05}  & {0.05}  & \\
     & \hbox{[Zn/Fe]}   &  -0.08  & +0.06 & +0.00 & +0.10  & \pm{0.05}  & {0.11} & \\
\hline
\noalign{\smallskip}
\hbox{} & & \hbox{$\Delta$T$_{\rm eff}$} & \hbox{$\Delta$$\log$ g}
& \hbox{$\Delta$v$_{t}$} 
& \hbox{($\sum$x$^{2}$)$^{1/2}$ }  & \hbox{continuum} & \hbox{($\sum$x$^{2}$)$^{1/2}$ } \\
& & \hbox{(+200 K)} & \hbox{(+0.4)} & \hbox{(+0.2 kms$^{-1}$})& {\rm (parameters)} &
 &    {\rm (final)} &\\
\noalign{\smallskip}
\hline
\hbox{GIRAFFE} &\hbox{[C/Fe](CI)} &  +0.00  & +0.00 & +0.00 & +0.00 & \pm{0.05}  & {0.05}  & \\
\hbox{BW-b6} &\hbox{[C/Fe](CH)}    &  +0.00  & +0.00 & +0.00 & +0.00 & \pm{0.05}  & {0.05} & \\
     &\hbox{[N/Fe]}        &  -0.10  & +0.05 & +0.00 & +0.11 & \pm{0.02}  & {0.07} & \\
     & \hbox{[O/Fe]}       &  +0.00  & +0.05 & +0.00 & +0.05 & \pm{0.05}  & {0.07}  & \\
     & \hbox{[Zn/Fe]}      &  -0.10  & +0.12 & +0.00 & +0.16 & \pm{0.05}  & {0.17} & \\
\noalign{\hrule\vskip 0.1cm}
& &  \hbox{$\Delta$T$_{\rm eff}$} & \hbox{$\Delta$$\log$ g}
& \hbox{$\Delta$v$_{t}$} 
& \hbox{($\sum$x$^{2}$)$^{1/2}$ }  & \hbox{continuum} & \hbox{($\sum$x$^{2}$)$^{1/2}$ } \\
& & \hbox{(-150 K)} & \hbox{(+0.2)} &  \hbox{(+0.1 kms$^{-1}$})& {\rm (parameters)} &
 &    {\rm (final)} &\\
\noalign{\smallskip}
\hline
\noalign{\smallskip}
\hbox{UVES} &\hbox{[C/Fe](CI)} &  +0.08  & +0.00 & +0.00 & +0.08  & \pm{0.02}  & {0.08}  & \\
\hbox{B6-b3}&\hbox{[C/Fe](CH)} &  +0.00  & +0.00 & +0.00 & +0.00  & \pm{0.05}  & {0.05} & \\
     &\hbox{[N/Fe]}     &  +0.08  & +0.00 & +0.00 & +0.08  & \pm{0.02}  & {0.08} & \\
     & \hbox{[O/Fe]}    &  +0.00  & +0.03 & +0.00 & +0.03  & \pm{0.05}  & {0.06}  & \\
     & \hbox{[Zn/Fe]}   &  -0.05   & +0.03 & +0.00 & +0.06  & \pm{0.05}  & {0.08} & \\
\hline
\noalign{\smallskip}
\hbox{} & & \hbox{$\Delta$T$_{\rm eff}$} & \hbox{$\Delta$$\log$ g}
& \hbox{$\Delta$v$_{t}$} 
& \hbox{($\sum$x$^{2}$)$^{1/2}$ }  & \hbox{continuum} & \hbox{($\sum$x$^{2}$)$^{1/2}$ } \\
& & \hbox{(-200 K)} & \hbox{(+0.4)} & \hbox{(+0.2 kms$^{-1}$})& {\rm (parameters)} &
 &    {\rm (final)} &\\
\noalign{\smallskip}
\hline
GIRAFFE &\hbox{[C/Fe](CI)} &  +0.10  & +0.00 & +0.00 & +0.10 & \pm{0.05}  & {0.11}  & \\
B6-b3 &\hbox{[C/Fe](CH)}    &  +0.00  & +0.00 & +0.00 & +0.00 & \pm{0.05}  & {0.05} & \\
     &\hbox{[N/Fe]}        &  +0.10  & +0.00 & +0.00 & +0.10 & \pm{0.02}  & {0.10} & \\
     & \hbox{[O/Fe]}       &  +0.00  & +0.05 & +0.00 & +0.05 & \pm{0.05}  & {0.07}  & \\
     & \hbox{[Zn/Fe]}      &  -0.07  & +0.06 & +0.00 & +0.09 & \pm{0.05}  & {0.10} & \\
\noalign{\vskip 0.1cm}
\hline\hline
\noalign{\smallskip}
\end{array}
\]
\end{table*}

\section{Results}
\label{results-discussion}

Table \ref{results} reports the stellar parameters 
  by Zoccali et al. (2008) for the GIRAFFE sample.
For stars for which we have both UVES and GIRAFFE
spectra, the two sets of parameters and results are reported in Table B.1, with
the UVES ones first, marked with a star (*), and the GIRAFFE ones just below.
In this Table are given the OGLE, GIRAFFE and UVES names, stellar parameters,
the derived N, O and Zn abundances, and the $\alpha$-elements Mg, Si, Ca, and Ti
analysed by Gonzalez et al. (2011). 

\subsection{Oxygen abundances}

In FB17 we discussed  the available
previous work on  bulge samples with reported
derivations of oxygen abundances. These were  
the bulge dwarfs by Bensby et al. (2013), the red
giant stars from Alves-Brito et al. (2010)
that were carried out in the optical for the same stars as
 in Mel\'endez et al. (2008), 
Cunha \& Smith (2006), Ryde et al. (2010), Rich et al. (2012),
Johnson et al. (2014), Rich \& Origlia (2005) and
Fulbright et al. (2007).

 We now compare the present oxygen abundances for the GIRAFFE sample
 together with those from the UVES sample, given in FB17, compared with:
 a) the reanalysis of stellar parameters
 carried out by J\"onsson et al. (2017) for 23 stars of the
same UVES data, for which they derived oxygen abundances,
except for one of them (B3-f1) that FB17 did not
include in their study;
 b) Ryde et al. (2010) where five stars of our UVES sample were
included, plus another six stars;
c) recent results by Schultheis et al. (2017), where comparisons
with a fraction of the present stars were given;
d) recent results for microlensed dwarf stars by Bensby et al. (2017).

Figure \ref{ofe} shows the [O/Fe] vs. [Fe/H] 
for the present sample (excluding
six N-rich, O-poor stars), plotted together
with oxygen abundances from UVES data for stars in common, analysed
both by FB17, and J\"onsson et al. (2017), as well as
 Ryde et al. (2010), Schultheis et al. (2017), and Bensby et al. (2017).
Also included are  
 recent oxygen abundances for metal-poor stars located in
outer bulge fields:
five stars from  Garc\'{\i}a-Perez et al. (2013), two stars from
Howes et al. (2016), and three stars from Lamb et al. (2017).

In Figures \ref{ofe} and \ref{znfe} we overplot the
 behaviour of oxygen and zinc respectively, in
chemodynamical models  representing a classical bulge, 
as described in FB17,
and briefly summarized as follows.
The evolution of the model was followed up to 13 Gyr,
and although the bulge is formed rapidly
the star formation goes on, and the stellar mass is built up during
 at least $\approx 3$ Gyr, allowing for
a contribution from type Ia supernovae (SNIa).
The best fit model for oxygen, based on previous
data, was  assumed to have a specific star formation rate of 
$\nu_{\rm SF}=$ 0.5 Gyr$^{-1}$, following conclusions by
FB17, and Cavichia et al. (2014).

{\it Comparison with literature}

 Part of the present data set has been under study recently
 by J\"onsson et al. (2017) and Schultheis et al. (2017). 
Given that this may be considered as
a reference sample for bulge studies, it is important to compare these different
analyses.

In Table \ref{jonsson} we give the stellar parameters rederived by
J\"onsson et al. (2017), their oxygen abundances given in
$\epsilon$(O)\footnote{$\epsilon$(X) = log(n(X)/n(H))+12, where n = number 
density of atoms, is a standard notation}, and their [O/Fe] =
$\epsilon$(O)$_{*}$ - $\epsilon$(O)$_{\odot}$  - [Fe/H], assuming
  $\epsilon$(O)$_{\odot}$ = 8.76 (Steffen et al. 2015).
For a comparison with the present work, the 
stellar parameters from Zoccali et al. (2006) adopted in the present
work  and in FB17 are reported in the same table,
 and in the last column the abundance ratio
of oxygen-to-iron as rederived by FB17.

\begin{figure}
\centering
\includegraphics[width=\columnwidth]{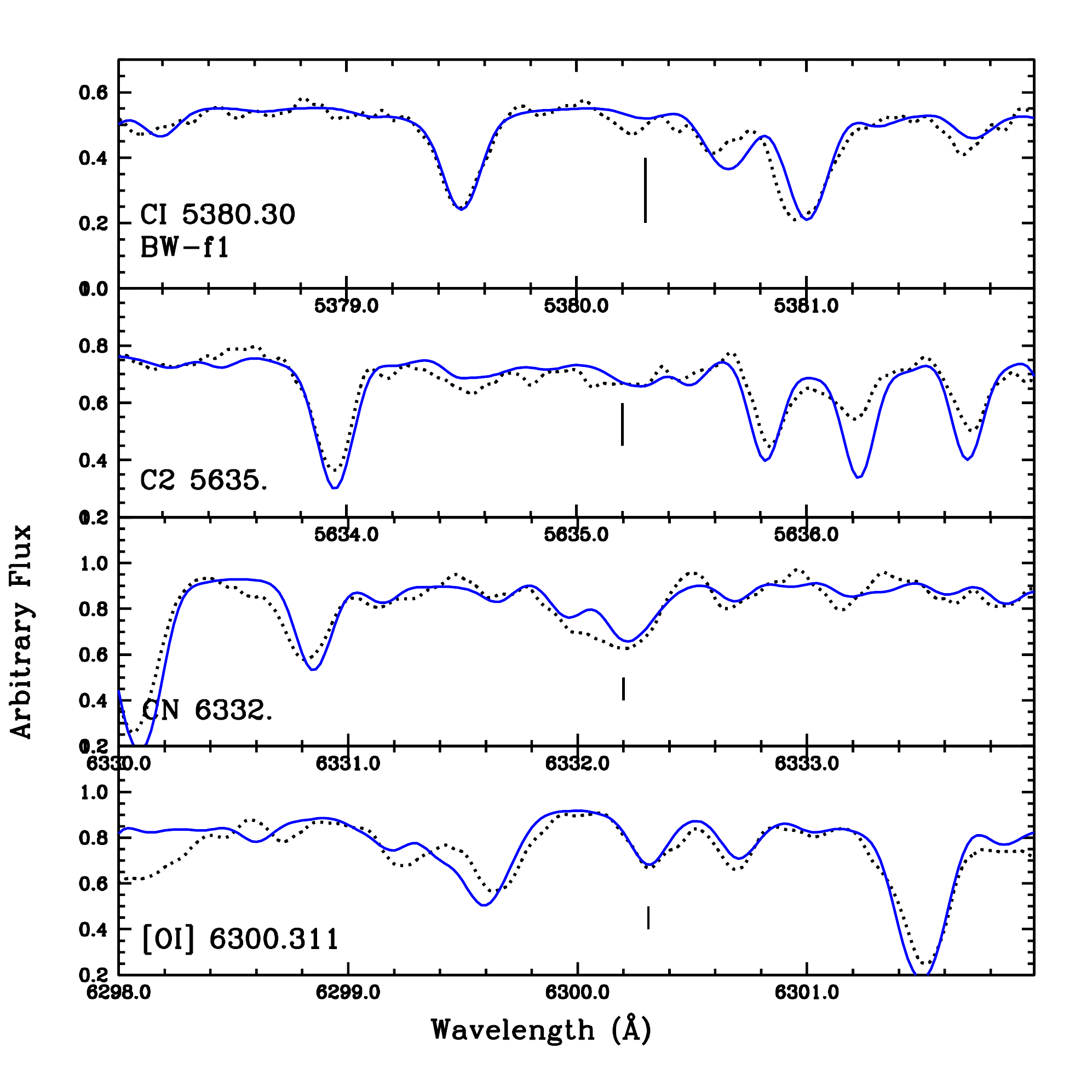}
\includegraphics[width=\columnwidth]{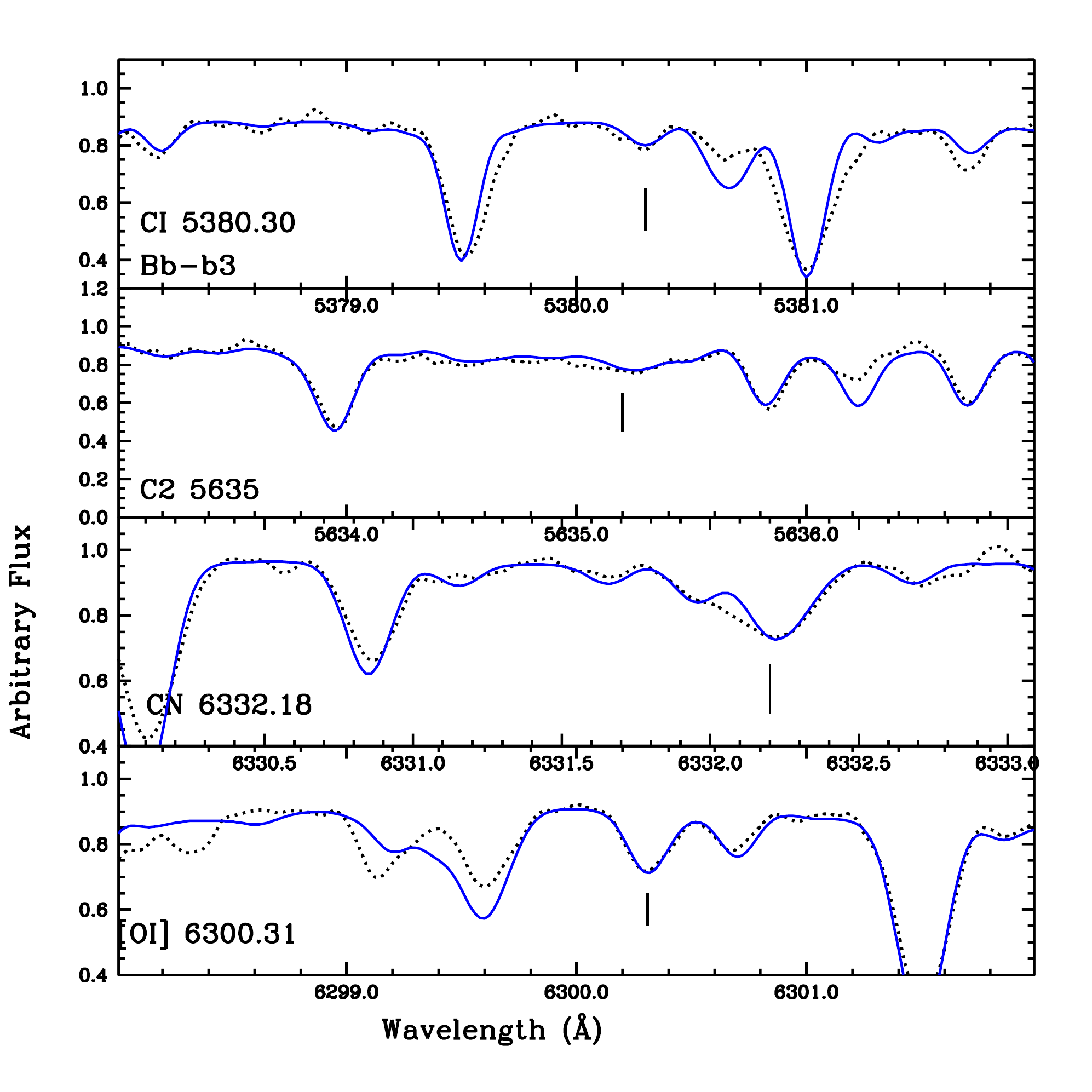}
\includegraphics[width=\columnwidth]{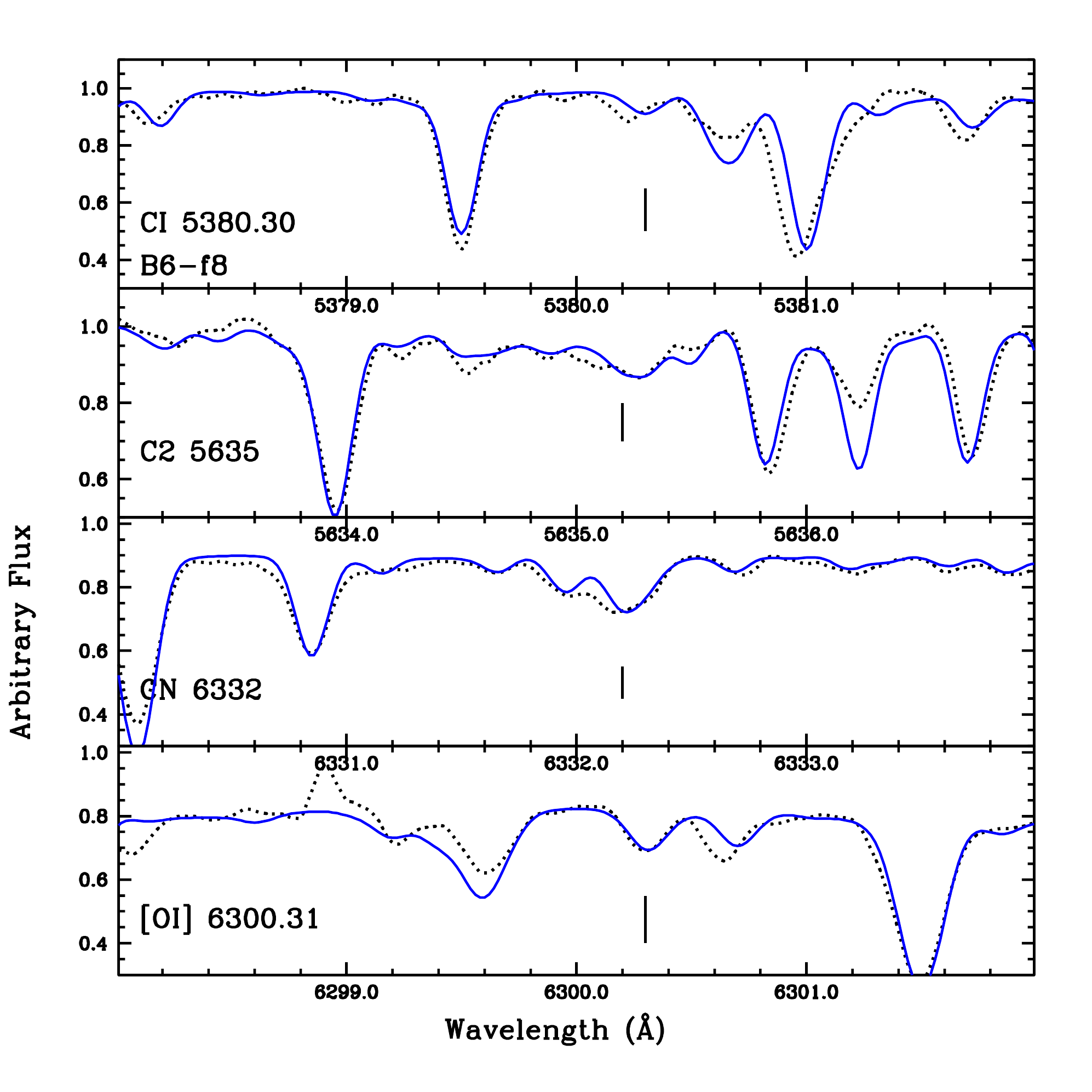}
\caption{\label{bwf1} CNO abundances rederived for
stars BW-f1 B6-b3, B6-f8, adopting stellar parameters defined by
J\"onsson et al. (2017).  Symbols: black dotted line: observed spectra;
blue solid line: synthetic spectra. }
\end{figure}

We restrict these comparisons to the BW
and -6$^{\circ}$ samples studied in the present work.
For three stars (B6-b3, B6-f3, and B6-f8) the effective temperatures
differ by $\Delta$T$_{\rm eff}$(Zoccali+06-J\"onsson+17) = -237 K,
+364, and +232 K.
For three stars the [O/Fe] value is different by more than 0.2 dex,
with [O/Fe](J\"onsson+17,FB17):
BW-f1: +0.45, -0.18; B6-b3: +0.13, -0.12; B6-f8: +0.03, -0.20,
and we inspect these stars in particular more closely.

For these three metal-rich stars: BW-f1, B6-b3, and B6-f8, we 
employed the new stellar parameters from J\"onsson et al. (2017), and
rederived the
C,N, and O abundances in the same way described in FB17,
and the results are reported in Table \ref{jonsson}.
 Only for BW-f1 the oxygen
abundance differs from that of J\"onsson et al., whereas for the other
two stars they are similar.
Whereas the [O/Fe] values are comparable,  
it seems to us that both sets of parameters may be hinting at
 uncertainties:
on the one hand, for some cases  the
gravities may be too high in J\"onsson et al.
 given that we are dealing with
 stars located one magnitude above the horizontal branch, 
and on the other, the Zoccali et al. metallicities for 
some of the metal-rich stars
may be too high.
In the mean $\Delta$[Fe/H](Zoccali+06-J\"onsson+17)$\sim$0.05dex.
Except for a large difference in [O/Fe] for BW-f1,
the two sets of results agree rather well, and are well-reproduced by
the models.

The uncertainty is made more clear if we compare the parameters 
of J\"onsson et al. (2017), and
 APOGEE data from Schultheis et al. (2017), with respect
to Zoccali et al. (2006, 2008). These differences could
be taken as the uncertainty expected from different analyses.
 The comparison of parameters results in the following differences:
in effective temperatures 
$\Delta$T$_{\rm eff}$\-(J\"onsson+17\--\-Zoccali+06)\-=\--94 K,
and  $\Delta$T$_{\rm eff}$\-(Schultheis+17\--Zoccali+08)\-=\-+250 K;
and in gravity values 
 $\Delta$log~g\-(J\"onsson+17\--Zoccali+06)\-=\-+0.46 and 
 $\Delta$log~g\-(Schultheis+17-\-Zoccali+08)\-=\-+0.10 (excluding the very
discrepant star 2MASS 18042724-3001108).
 Schultheis et al. (2017) also found $\Delta$[Fe/H]
(Schultheis+17\--Zoccali+08) = 0.1 dex,  and it is different
if considering only the metal-poor and metal-rich stars separately 
where stars with [M/H]$<$0 are systematically more metal-poor in
Zoccali et al. (2008) with respect to the APOGEE 
measurements.  The differences are larger in effective temperatures 
with respect to Schultheis et al. and in gravity with respect to 
J\"onsson et al. (2017). The results will become more accurate in the near future, due to the possibility of fixing
gravity values with data from the next release of the
 Gaia collaboration (2017).

\begin{table*}
\caption{Sample of stars observed with both FLAMES-UVES, and reanalysed by
J\"onsson et al. (2017).
Columns 8-11: stellar parameters from Zoccali et al. (2006); Column 12:
[O/Fe] abundances from Fria\c ca \& Barbuy (2016).}             
\label{jonsson} 
\scalefont{1.0}     
\centering          
\begin{tabular}{lrrrrrrrrrrrrrrrrr}     
\noalign{\vskip 0.1cm}
\noalign{\hrule\vskip 0.1cm}
\noalign{\vskip 0.1cm}    
 & \multicolumn{6}{c}{\hbox{\bf J\"onsson+17}} 
&\hbox{ New C,N,O} & \multicolumn{4}{c}{\hbox{\bf Zoccali+06}} & FB17 \\
\noalign{\vskip 0.1cm} 
\hline\hline    
\noalign{\smallskip}
 Star   & T$_{\rm eff}$ & log~g & [Fe/H]& v$_{\rm t}$& $\epsilon_{*}$(O) & [O/Fe] 
& C,N,O &
 T$_{\rm eff}$ & log~g & [Fe/H]& v$_{\rm t}$& [O/Fe] & \\
\hline    
\noalign{\smallskip}
 B3-b1 & 4414 & 1.35 & -0.92 &  1.41 &  8.22 & +0.38&---& 4300 &1.7 & -0.78& 1.5&  +0.35& \\
 B3-b5 & 4425 & 2.70 &  0.22 &  1.43 &  8.87 & -0.11&---& 4600 &2.0 &  0.11& 1.5&  -0.30& \\
 B3-b7 & 4303 & 2.36 &  0.05 &  1.58 &  8.80 & -0.01&---& 4400 &1.9 &  0.20& 1.3&  -0.20& \\
 B3-b8 & 4287 & 1.79 & -0.70 &  1.46 &  8.47 & +0.41&---& 4400 &1.8 & -0.62& 1.4&   0.30& \\
 B3-f1 & 4485 & 2.25 & -0.18 &  1.88 &  8.74 & +0.16&---& 4500 &1.9 &  0.04& 1.6&   0.10& \\
 B3-f3 & 4637 & 2.96 &  0.21 &  1.89 &  8.98 & +0.01&---& 4400 &1.9 &  0.06& 1.7&  -0.10& \\
 B3-f4 & 4319 & 2.60 & -0.15 &  1.50 &  8.77 & +0.16&---& 4400 &2.1 &  0.09& 1.5&   0.10& \\
 B3-f8 & 4436 & 2.88 &  0.21 &  1.54 &  8.79 & -0.18&---& 4800 &1.9 &  0.20& 1.5&  -0.30& \\
 BW-b6 & 4262 & 1.98 & -0.35 &  1.44 &  8.60 & +0.19&---& 4200 &1.7 & -0.25& 1.3&   0.15& \\
 BW-f1 & 4359 & 2.51 &  0.25 &  1.93 &  8.96 & +0.45 &-0.10,0.75,0.00& 4400 &1.8 &  0.32& 1.6&  -0.18& \\
 BW-f6 & 4117 & 1.43 & -0.46 &  1.69 &  8.55 & +0.25&---& 4100 &1.7 & -0.21& 1.5&   0.20& \\
 BW-f7 & 4592 & 2.96 &  0.53 &  1.50 &  9.10 & -0.19&---& 4400 &1.9 &  0.11& 1.7&  -0.25& \\
 B6-b3 & 4468 & 2.48 &  0.02 &  1.67 &  8.91 & +0.13&0.15,0.45,0.20& 4700 &2.0 &  0.10& 1.6&  -0.12& \\
 B6-b4 & 4215 & 1.38 & -0.65 &  1.68 &  8.43 & +0.32&---& 4400 &1.9 & -0.41& 1.7&   0.30& \\
 B6-b5 & 4340 & 2.02 & -0.51 &  1.34 &  8.49 & +0.24&---& 4600 &1.9 & -0.37& 1.8&   0.15& \\
 B6-b6 & 4396 & 2.37 &  0.16 &  1.77 &  8.86 & -0.15&---& 4600 &1.9 &  0.11& 1.8&  -0.10& \\
 B6-b8 & 4021 & 1.90 &  0.03 &  1.45 &  8.68 & -0.11&---& 4100 &1.6 &  0.03& 1.3&  -0.03& \\
 B6-f1 & 4149 & 2.01 &  0.07 &  1.65 &  8.84 & +0.01&---& 4200 &1.6 & -0.01& 1.5&   0.03& \\
 B6-f3 & 4565 & 2.60 & -0.38 &  1.28 &  8.63 & +0.23&---& 4800 &1.9 & -0.29& 1.3&   0.15& \\
 B6-f8 & 4470 & 2.78 &  0.10 &  1.30 &  8.89 & +0.03&0.10,0.25,0.10& 4900 &1.8 &  0.04& 1.6&  -0.20& \\
\hline                   
\hline                  
\end{tabular}
\end{table*}

\subsection{Zinc abundances} 
\label{zincabundances}

Figure \ref{znfe} gives [Zn/Fe] vs. [Fe/H] for the sample stars, 
together with the UVES sample from Barbuy et al. (2015),
the recent Zn abundances derived for 90 microlensed bulge dwarf stars
by Bensby et al. (2017),  and 
metal-poor stars analysed by Howes (2015a), Howes et al. (2014, 2015b, 2016)
 and Casey \& Schlaufman (2016).
 This figure shows that for bulge metal-poor stars
with [Fe/H]$\simless$-1.4, Zn is enhanced with [Zn/Fe] $\sim$ +0.4.
This behaviour is in agreement with the same trend of
 increasing [Zn/Fe] values with decreasing metallicities
for thick disk and halo stars as shown in Fig. 7 by Barbuy et al.
(2015), where results by Bensby et al. (2014), 
 Ishigaki et al. (2013), Nissen \& Schuster (2011),
Mishenina et al. (2011), Prochaska et al. (2000), Reddy et al. (2006),
and Cayrel et al. (2004) were reported.

In Fig. \ref{alpha},  Zn abundances are plotted,
compared with the $\alpha$-element abundances of
O, as derived in the present work,
and Mg, Si, Ca, and Ti from Gonzalez et al. (2011).
The trend shown by Zn appears similar to that of the $\alpha$-elements,
and more closely to oxygen, silicon, and calcium. 
The low [Zn/Fe] for high metallicity
stars is compatible with the oxygen abundances.

Chemodynamical evolution models of zinc
were computed for a small classical spheroid,
with a baryonic mass of 2$\times$10$^9$ M$_{\odot}$,
and a dark halo mass $M_{H}$= 1.3$\times$10$^{10}$ M$_{\odot}$,
by Barbuy et al. (2015), FB17.
The code allows for inflow and outflow of gas, treated
with hydrodynamical equations coupled with chemical evolution.

As discussed in Barbuy et al. (2015), the yields
 from core-collapse SN II by  WW95 underestimate
the Zn abundance at low metallicities.
Hypernovae, as defined by Nomoto et al. (2006, 2013), Umeda \& Nomoto
(2002, 2003, 2005), and Kobayashi et al. (2006), 
reproduce better the enhanced zinc-to-iron abundances in metal-poor
stars. There are certain differences with respect to
those models in the present work.
In Barbuy et al. (2015) the contribution of hypernovae was
included for metallicities Z/Z$_{\odot}$$\leq$0.0001, which reproduced
well the abundances of DLAs at low metallicities.
In the present work, the chemical evolution calculations
took into account the
core-collapse SN II models of WW95, for metallicities
 Z/Z$_{\odot}$$>$0.01, and for  Z/Z$_{\odot}$$<$0.01
we used a weighted mean of WW95 and
 the hypernovae yields by Kobayashi et al. (2006),
 fitting well the data below [Fe/H]$\simless$-1.6.
There is still an unsolved gap at the moderate metallicities
of -1.6$<$[Fe/H]$<$-0.9.
 There is therefore a mismatch between the models and
the data in this metallicity range.
It is important to note that chemodynamical models are suitable
to indicate the inflexion of the [X/Fe] values due to
enrichment of Fe from SNIa.

\subsubsection{Comparison with literature}
\label{literaturezn}

Comparisons with literature 
Zn abundances of microlensed dwarf bulge stars
 by Bensby et al. (2013), were discussed in Barbuy et al. (2015).
 In Fig. \ref{znfe} we show the updated abundances for microlensed 
bulge dwarfs by Bensby et al. (2017).
There is good agreement between the present
results and Barbuy et al. (2015)
and those by Bensby et al. at metallicities $-$1.4 $<$ [Fe/H] $<$ 0.0,
whereas the behaviour for metal-rich giants with [Fe/H] $>$ 0.0 
are distinct. The microlensed dwarfs show a contant [Zn/Fe],
whereas the bulge red giants show a decreasing trend
with metallicity,  although with 
a large spread of $-$0.6 $<$ [Zn/Fe] $<$ +0.15.

 At the high metallicity end, since there is progressive enrichment in Fe
by SNIa, a constant [Zn/Fe] would imply 
that there is chemical enrichment in both Zn and Fe
 on similar timescales.
Instead, a decrease of [Zn/Fe] would correspond
to the enrichment in Fe by SNIa, with no enrichment in Zn
by the same SNIa, as happens for the $\alpha$-elements.

 This discrepancy has been addressed by Duffau et al. (2017),
who found, at supersolar metallicities, a decreasing
[Zn/Fe] for red giants, and constant [Zn/Fe] for dwarfs.
 Their interpretation is that the dwarfs are old and
the red giants are young. This interpretation cannot
be applied here, given that at least part of the bulge
metal-rich red giant stars should be old, as can be seen
in the distribution of ages given in
 Bensby et al. (2017, see their Figs. 14 and 15).

The derivation of [Zn/Fe] in stars of dwarf galaxies by
 Sk\'ulad\'ottir et al. (2017, 2018, and references therein)
indicated a decreasing [Zn/Fe]  with increasing metallicities.
This behaviour is in agreement with an Fe enrichment by SNIa, but not
with a Zn enrichment.

\begin{figure*}
\centering
\includegraphics[width=\columnwidth, angle=-90]{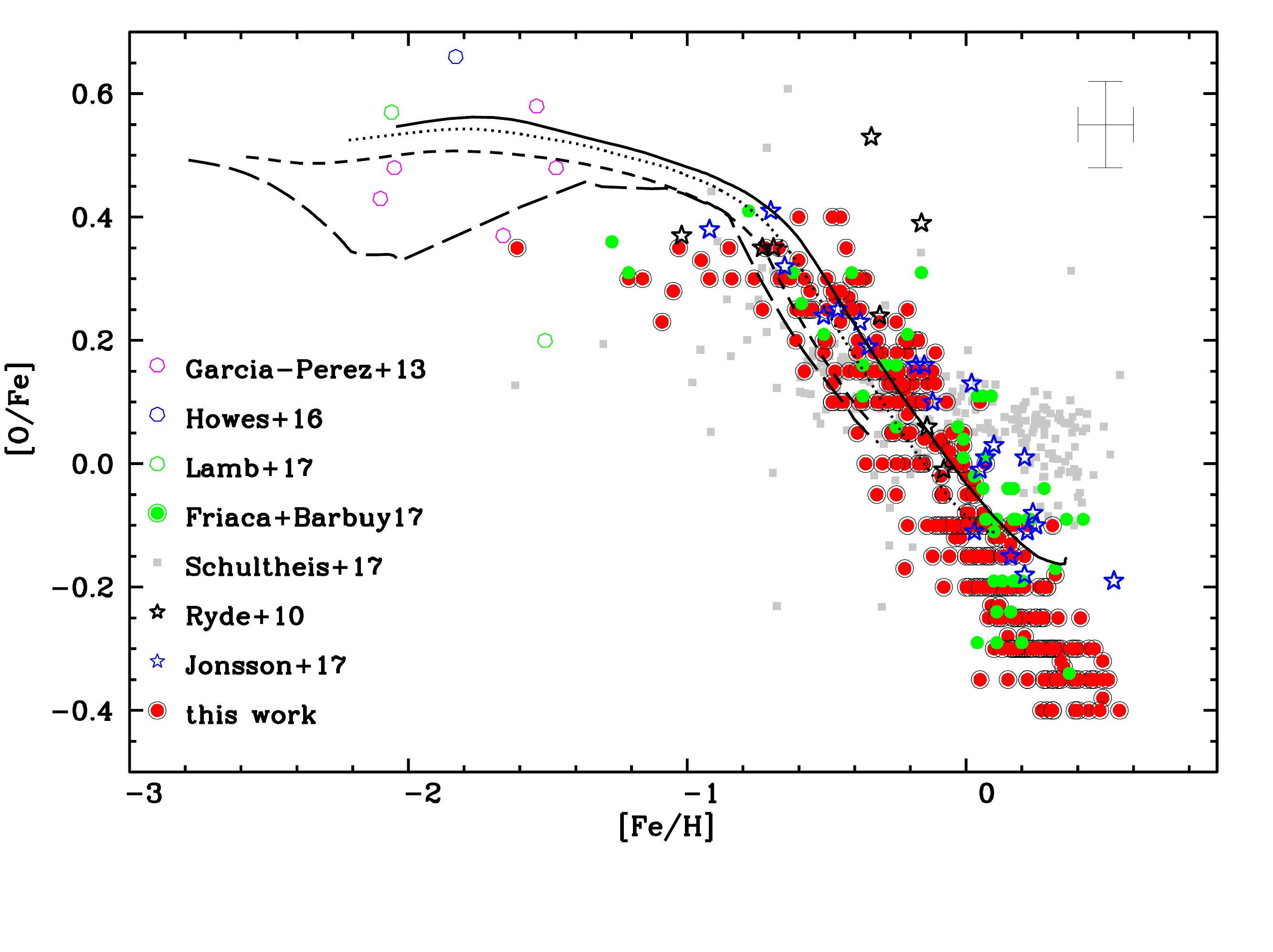}
\caption{[O/Fe] vs. [Fe/H]  for 351 red giants (excluding N-rich, O-poor ones).
Chemodynamical evolution models from FB17 with formation timescale of 2 Gyr, 
or specific star formation rate of 0.5 Gyr$^{-1}$ are overplotted.
Solid lines:  5$<$0.5 kpc; dotted lines: 0.5$<$r$<$1 kpc;
dashed lines: 1$<$r$<$2 kpc; long-dashed lines: 2$<$r$<$3 kpc.
 Symbols:
red filled circles: present work;
green filled circles: Fria\c ca \& Barbuy (2017); 
blue stars: J\"onsson et al. (2017); 
black stars: Ryde et al. (2010);
grey filled squares: Schultheis et al. (2017);
magenta open circles: Garc\'{\i}a-P\'erez et al. (2013);
blue open circles: Howes et al. (2016); 
green open circles: Lamb et al. (2017).
Errors indicated correspond to 0.1dex in both [Fe/H] and [O/Fe].
The model lines correspond to different radii from the Galactic
centre:
solid lines:  r$<$0.5 kpc; dotted lines: 0.5$<$r$<$1 kpc;
dashed lines: 1$<$r$<$2 kpc; long-dashed lines: 2$<$r$<$3 kpc.
A typical error bar is indicated in the right upper corner.}

\label{ofe} 
\end{figure*}

\begin{figure*}
\centering
\includegraphics[width=\columnwidth, angle=-90]{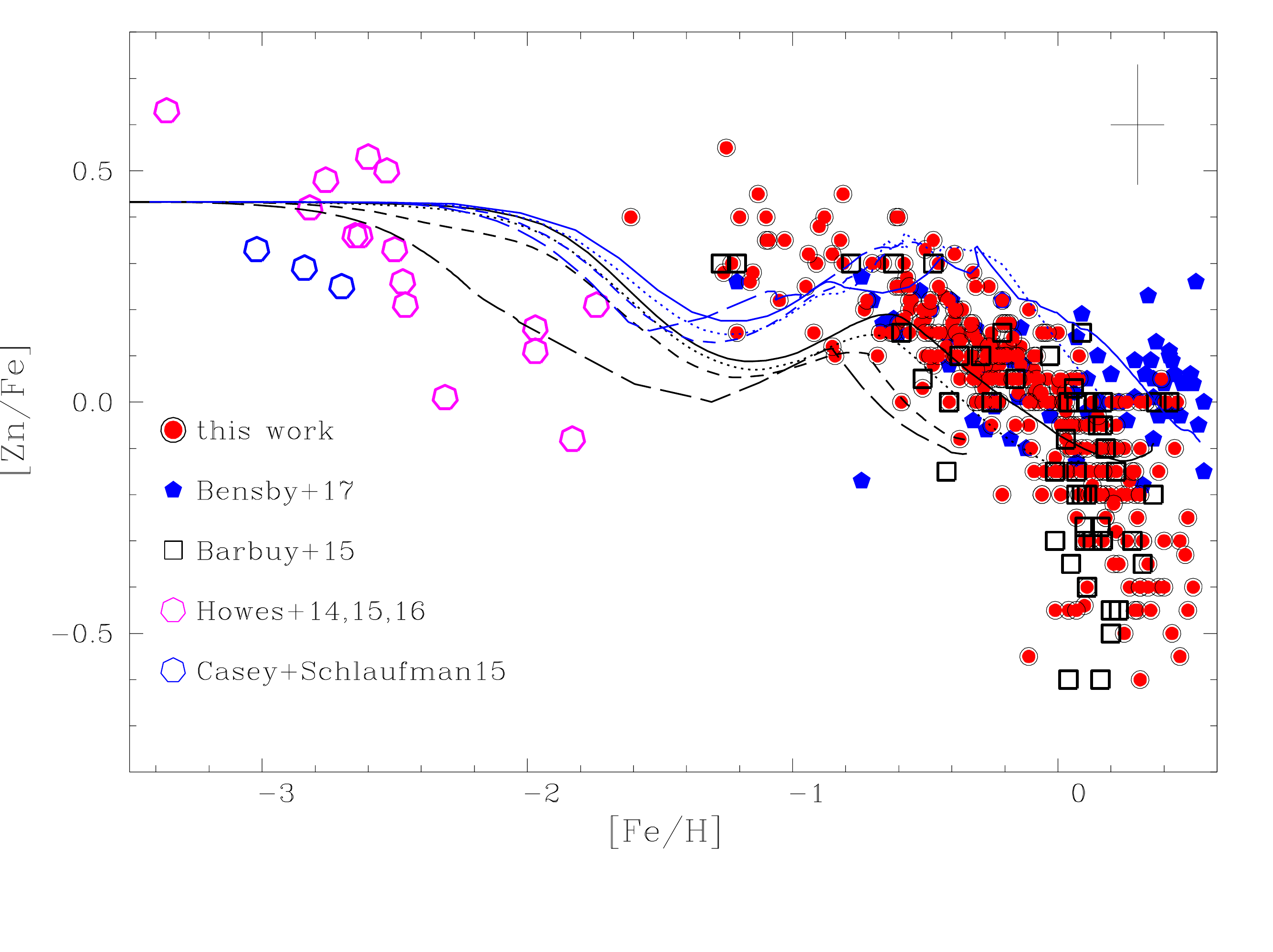}
\caption{[Zn/Fe] vs. [Fe/H] for the present sample (333 stars),
compared with literature.
Symbols: 
red filled circles: present work;
black squares: results for stars in common
 based on UVES data (Barbuy et al. 2015);
blue filled pentagons: Bensby et al. (2017);
magenta open heptagons: Howes et al. (2015, 2016);
blue open heptagons: Casey \& Schlaufman (2015).
Chemodynamical evolution models by FB17 with
formation timescale of 2 (black lines) and 3 Gyr (blue lines), 
or specific star formation rate of 0.5, 0.3 Gyr$^{-1}$ are overplotted.
 The model lines correspond to different radii from the Galactic
centre:
solid lines:  r$<$0.5 kpc; dotted lines: 0.5$<$r$<$1 kpc;
dashed lines: 1$<$r$<$2 kpc; long-dashed lines: 2$<$r$<$3 kpc.
A typical error bar is indicated in the right upper corner, corresponding
to a mean between the two reference stars (Table \ref{errors2}).}
\label{znfe} 
\end{figure*}

\begin{figure*}
\centering
\includegraphics[width=\columnwidth]{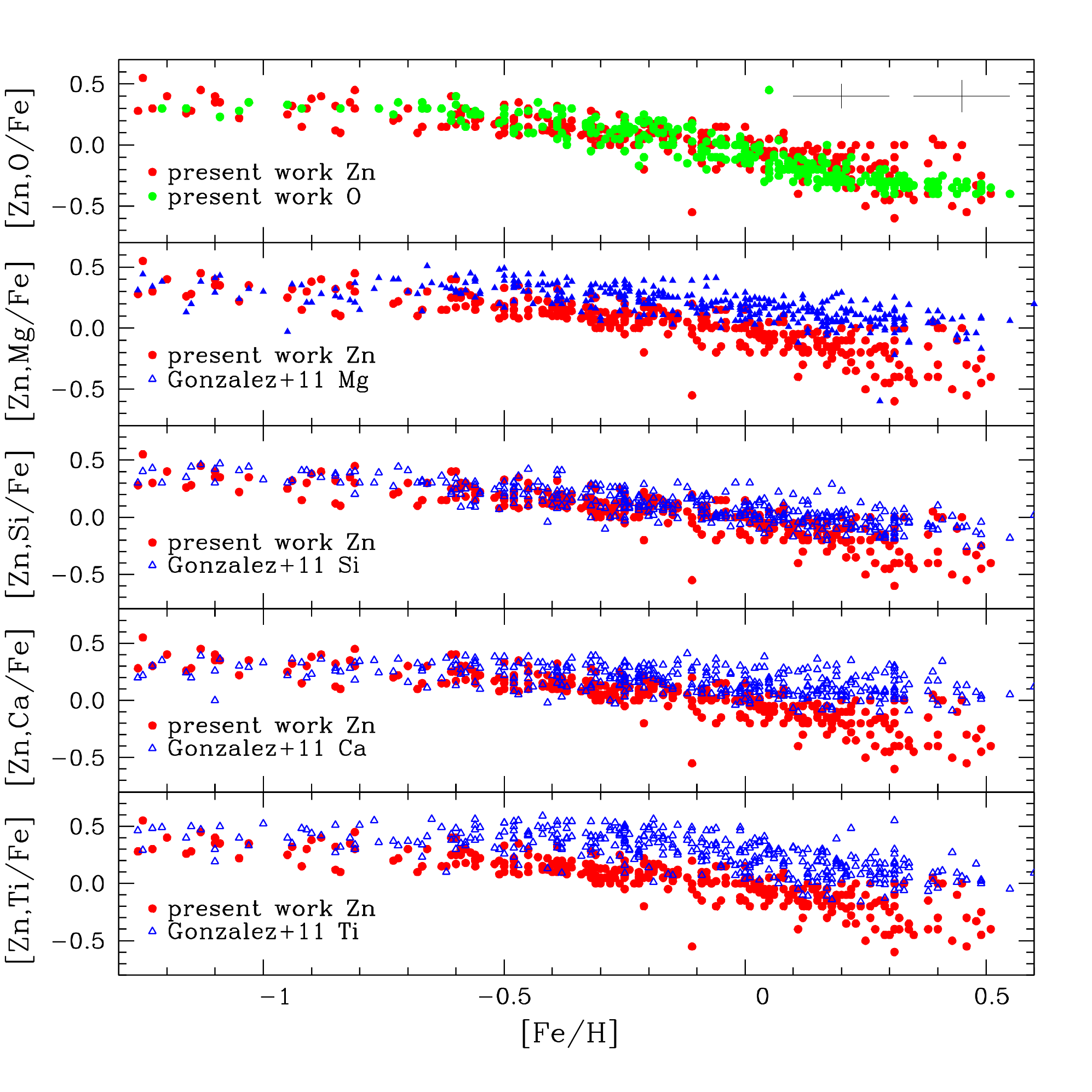}
\caption{[O, Mg, Si, Ca,Ti/Fe] vs. [Fe/H] and [Zn/Fe] vs. [Fe/H]
for the 417 red giants. Symbols: Red filled circles: O from this work;
Green filled circles: Zn from this work; Blue open triangles: 
Mg, Si, Ca, and Ti from Gonzalez et al. (2011). Typical error bars are
indicated for [$\alpha$/Fe] and [Zn/Fe]. }
\label{alpha} 
\end{figure*}

\subsubsection{Comparison with Damped Lyman-alpha systems}

\begin{figure*}
\centering
\includegraphics[width=\columnwidth]{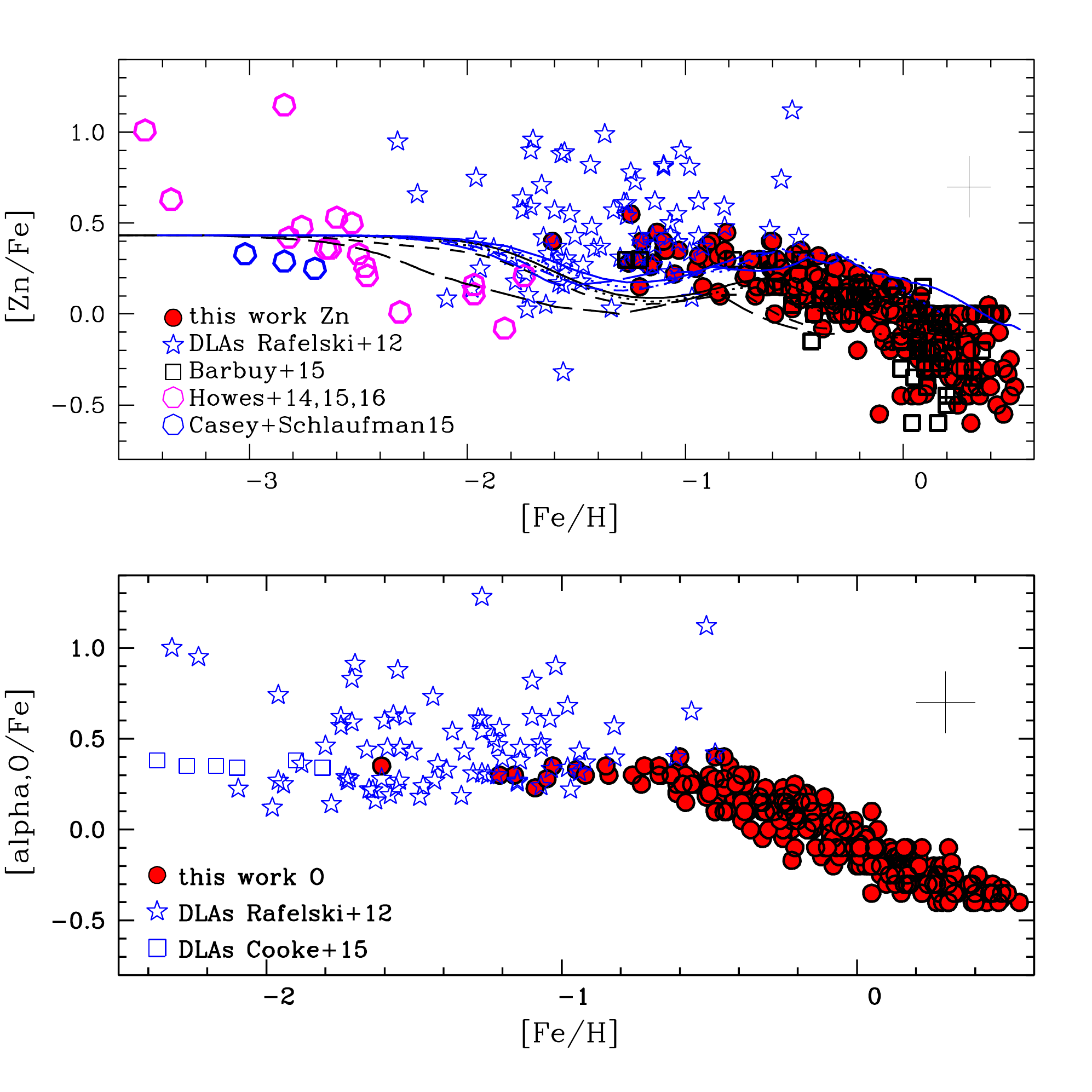}
\caption{a) [Zn/Fe] vs. [Fe/H]: same as in Fig. \ref{znfe}, including
the Damped Lyman-alpha systems data by Rafelski et al. (2012).
Models from FB17 are overplotted (same details as in Fig. \ref{znfe});
b)  [$\alpha$/Fe] vs. [Fe/H]: data by Rafelski et al. (2012),
[O/Fe] in DLAs by Cooke et al. (2015) and present results for [O/Fe]
in the sample stars. 
The model lines  in panel a) correspond to
the same Galactic bulge radii as in Fig. \ref{znfe}.
A typical error bar is indicated in the right upper corner
in both panels: for [Zn/Fe] it corresponds to a mean value of the
two reference stars given in Table \ref{errors2}. For [$\alpha$/Fe] the
error is of $\pm$0.10 for both the stars and DLAs.}
\label{znfeDLA} 
\end{figure*}

Comparisons of Zn abundances with data from Akerman et al. (2005), Cooke et al. (2013), and Vladilo et al. (2011) were shown in Barbuy et al. (2015). Using careful dust corrections, Barbuy et al. (2015) concluded that the DLAs fall into the same region of [Zn/Fe] vs. [Fe/H] as thick disk and bulge stars. On the other hand, a comparison of the metallicity of DLAs to thick disk stars in Rafelski et al. (2012) showed that while there is some overlap, the median DLA population is more metal poor than the thick disk stars. While Rafelski et al. (2012) did not apply dust corrections, metallicities were determined using primarily [Si/H] and [S/H], which are less sensitive to dust than [Zn/H]. To investigate this further, we compared the [Zn/Fe] vs. [Fe/H] from our present work to the DLA data presented in Rafelski et al. (2012) in Figure  \ref{znfeDLA}. 
These data include a compilation of previous DLA systems selected to be unbiased with regard to their metallicities, including those from Akerman et al. (2005) and a subset of Vladilo et al. (2011). We note that the comparison in 
Figure  \ref{znfeDLA} must be taken with caution, because there are potential biases with the [Fe/Zn] values that are difficult to control. In the metal rich regime, Fe is strongly depleted by dust, while on the metal-poor side,
the oscillator strengths of Zn result in the absorption lines too weak to be detected in low-metallicity systems.
To reduce the biases from dust depletion and undetected Zn absorption lines, we limited our comparison in Figure  \ref{znfeDLA} to systems with
 $-$2.5$<$[alpha/H]$<-$1.0. In this comparison, 
no correction for dust is applied. 

Figure  \ref{znfeDLA}a shows an enhanced zinc-to-iron ratio for the DLA data which is consistent with the present sample, although DLAs typically reside at lower metallicities. We note that Figure  \ref{znfeDLA} exaggerates the difference in metallicity ([Fe/H]) due to the removal of higher metallicity systems to avoid biases caused by dust depletion of Fe. Other literature data similarly show a spread in [Zn/Fe] (Akerman et al. 2005; Cooke et al. 2013 - see Barbuy et al. 2015), but is also compatible with a [Zn/Fe] enhancement.  
 Cooke et al. (2015) argue instead that [Zn/Fe] in DLAs
can be assumed to drop to solar at [Fe/H]$\approx$$-$2.0,
 based on a compilation of halo stars data by Saito et al. (2009).
They assume therefore that Zn tracks Fe for [Fe/H]$>$$-$2.0.
However, we show in Figure  \ref{znfeDLA}a that both the present sample
 and the DLAs have elevated [Zn/Fe] at $-$2.5$<$[$\alpha$/H]$<$$-$1.0. 
Moreover, in Rafelski et al. (2012), we find that Zn and S trace each other
 one-to-one, not consistent with the solar value, 
but rather consistent with the models in Fenner et al. (2004),
suggesting that Zn behaves like an $\alpha$-element in DLAs,
meaning that it is enhanced in metal-poor DLAs. In conclusion, Zn and
$\alpha$-elements show similar behaviour in metal-poor DLAs,
and so can be expected to trace one another.

 Figures \ref{znfeDLA}a and b show a large scatter in the 
[Zn/Fe] and [$\alpha$/Fe] in DLAs, due to varying star formation 
histories of the galaxies hosting DLAs, and due to variations of
 dust depletion for different sightlines. There may also be
 variations due to the complexities in the way Zn is produced. 
 While Fe is produced in SNe Ia, Zn is produced 
in massive stars (WW95; Umeda \& Nomoto 2002). 
The value of [Zn/Fe] in DLAs therefore likely depends on both the star 
formation histories of the host galaxies (Fenner et al. 2004) and 
on possible dust depletion in Fe for any individual sightline. 
Therefore a  complementary investigation can be accomplished 
by studies of the $\alpha$-enhancement [$\alpha$/Fe] at 
[$\alpha$/H]$\simless$$-$1.0.

Figure  \ref{znfeDLA}b shows an $\alpha$-element enhancement of DLAs at 
[Fe/H]$<$$-$1.0 compared with [O/Fe] values for the present sample.
In Fig. \ref{znfeDLA}b we also
include [O/Fe] values derived for metal-poor DLAs by Cooke et al. (2015). 
The enhanced [$\alpha$/Fe] for stellar data with [Fe/H]$<$$-$0.6, 
is consistent with the alpha-element enhancement of the DLA data.

\section{Oxygen-poor, nitrogen-rich stars}
\label{nrich}

Enhanced nitrogen is expected in red giants due to
CN-cycle (Iben 1967), and extra-mixing (e.g. Smiljanic et al. 2009),
as reviewed by Karakas \& Lattanzio (2014, and references therein).
The situation is different for N-rich and O-poor stars,
which were first detected in globular clusters (e.g. Sneden et al. 1997).
These stars are not only O-poor and N-rich, but also Na-rich,
and anomalous also in Mg and Al.
In the case of bulge red giants,
Schiavon et al. (2017) identified N-rich stars, with a peak in metallicity
 at [Fe/H]$\sim$$-$1.0. They included in this category stars with
[N/Fe]$\simgreat$+0.5, which in their sample of 5140 bulge giants,
correspond to 58 of them, therefore in a proportion of 1.1\%.
Schiavon et al. interpreted these stars as second generation
members evaporated from globular clusters.
Carretta et al. (2009) have shown that second generation stars
have low O, and high N and Na. 

For this reason, for the
N-rich  as defined in Schiavon et al. (2017)
with [N/Fe]$\simgreat$+0.5 stars in our sample, we 
 also measured their Na abundances, using the
\ion{Na}{I} 6154.23 and  6160.75 {\rm \AA} lines,
adopting a hyperfine structure for total values of log gf = -1.56 and -1.26, 
 respectively.
The N-rich stars  fall in different cases, in terms of N, O: i)
N-rich and O-poor, but
for some of them we could not derive [O/Fe] due
to blends with telluric lines, and only a N-enhancement is 
reported;  ii) N-rich and O-normal
with [Fe/H]$\leq$-0.5; iii) one star very N-rich
 [N/Fe]$>$1.0 with [Fe/H]=+0.08 ([OI]
line is blended with telluric lines in this case).  
These selected stars
are listed in Table \ref{nrich}, where besides
the [Na/Fe] value reported,   [Mg/Fe] values are also given
for an indication of the $\alpha$-element enrichment in these stars
as compared with the oxygen abundances.

If the criterion of [N/Fe]$\geq$0.5 for stars with [Fe/H]$\leq$-0.5,
 is adopted, we find 21 stars,
corresponding to about 5\% of the sample.
If we consider the N-rich ones together with [Na/Fe]$>$0.0, 
then we have 3.5\% of them. Finally,
 if we discard the  N-rich but O-normal,
keeping only the O-poor ones ([O/Fe]$\simless$0.1),
then we have five stars left, corresponding to about 1\% of the sample,
in agreement with the percentage given by Schiavon et al. (2017).
It would be interesting to derive Al for these stars in order
to verify a possible Mg-Al anticorrelation also detected in second
generation globular cluster stars. The cause
of these anomalies is currently under debate in the literature,
with the more massive low-Z asymptotic giant branch stars 
as the likely site for such nucleosynthesis products (Renzini et al. 2015).

\begin{table}
\caption{N-rich and/or O-poor stars.
Na abundances are a mean of abundances from
\ion{Na}{I} 6154.23 and  6160.75 {\rm \AA} lines. }             
\label{nrich} 
\scalefont{1.0}     
\centering          
\begin{tabular}{lcccccccccccc}     
\noalign{\vskip 0.1cm}
\noalign{\hrule\vskip 0.1cm}
\hline\hline    
\noalign{\smallskip}
Star & [Fe/H]& [N/Fe] & [O/Fe] & [Mg/Fe]  & [Na/Fe] \\
\noalign{\smallskip}
\hline
\noalign {\vskip 0.05cm}
\noalign {\hskip 2.0cm N-rich/O-poor stars}
\noalign {\vskip 0.05cm}
\hline
bwb008 & -0.80 &  1.00 &  0.00 &  0.15 & +0.45  & \\
bwb122 & -0.81 &  0.70 & -0.05 &  0.21 & +0.25 & \\
bwb128 & -0.82 &  0.70 &  0.00 &  0.23 & +0.15 & \\
bwb130 & -0.85 &  0.70 &  0.10 &  0.26 & +0.15  & \\
b6b100 & -0.40 &  0.50 &  0.00 &  0.36 & +0.00 & \\
b6b011 & -1.13 &  1.00 &  ---  &  0.38 & +0.00 & \\ 
b6b016 & -0.81 &  0.70 &  ---  &  0.37 & $-$0.10 & \\
\hline
\noalign {\vskip 0.05cm}
\noalign {\hskip 2.0cm N-rich, O-normal stars}
\noalign {\vskip 0.05cm}
\hline
bwb087 & -1.21 & 0.70 & 0.30 & 0.38 & +0.20 & \\
bwb091 & -0.60 & 0.50 & 0.40 & 0.40 & $-$0.15  & \\
bwb093 & -0.67 & 0.80 & 0.30 & 0.15 & $-$0.30  & \\
bwb102 & -0.50 & 0.50 & 0.20 & 0.15 & $-$0.05  & \\
b6b009 & -1.03 & 0.50 & 0.35 & 0.35 & $-$0.30  & \\
b6b021 & -0.76 & 0.70 & 0.30 & 0.41 & +0.00 & \\
b6b024 & -1.16 & 0.50 & 0.30 & 0.26 & $-$0.20 & \\
b6b048 & -0.95 & 0.50 & 0.33 & 0.25 & $-$0.30 & \\
b6b062 & -0.60 & 0.60 & 0.33 & 0.17 & $-$0.05  & \\
b6b072 & -0.57 & 0.60 & 0.25 & 0.27 & +0.10  & \\
b6b077 & -0.84 & 0.50 & 0.30 & 0.10 & $-$0.30 & \\
b6b083 & -0.50 & 0.50 & 0.30 & 0.18 & +0.00 & \\
b6f037 & -0.51 & 0.50 & 0.18 & 0.20 & +0.30  & \\
\hline
\noalign {\vskip 0.05cm}
\noalign {\hskip 2.0cm  Very N-rich, high metallicity star}
\noalign {\vskip 0.05cm}
\hline
b6f015 & +0.08 & 1.10 &  ---  & --- & +0.10 & \\        
\hline                   
\hline                  
\end{tabular}
\end{table}

\section{Summary}

We studied oxygen and zinc abundances for 417 field red giants
in the Galactic bulge. We were able to derive Zn, O, and N abundances for 
333, 358 and 403 of them, respectively.
We have identified five stars, corresponding to a 1\%  of stars 
that are simultaneously N-rich ([N/Fe]$>$0.5), and O-poor
([O/Fe]$\simless$0.1), and this reduces to four stars if
 the more rigorous criterion of also being Na-rich 
([Na/Fe]$>$0.0) is applied.
According to Schiavon et al. (2017),
these characteristics could be attributed to
 evaporated second generation stars of globular clusters.

The sample contains a number of moderately metal-poor stars
(-1.7$<$[Fe/H]$<$-0.5) that define better the behaviour of
[O/Fe] and [Zn/Fe] vs.[Fe/H] in this metallicity range.
The present chemodynamical evolution modelling of a classical bulge
is able to reproduce the behaviour of O and Zn abundances in the 
Galactic bulge, except for Zn in the range 
$\sim$$-$1.6$\simless$[Fe/H]$\simless$$-$0.8, where  the yields from
WW95 show a drop. We remind the reader that the models presented here
 consider yields from WW95 for [Fe/H]$>$$-$2.0, and
a mean of models by WW95 and Kobayashi et al. (2006) for
$-$4$<$[Fe/H]$<-$2.

 The high [Zn/Fe] in very metal-poor stars favours enrichment from hypernovae,
as defined by Nomoto et al. (2013 and references therein)
acting at these low metallicities.
In Damped Lyman-alpha systems (DLAs), 
a high [Zn/Fe] in metal-poor DLAs is also well reproduced 
by hypernovae yields. In DLAs
Zn appears to behave similarly to $\alpha$ elements, 
and show an enhancement of [$\alpha$/Fe] similar to the metal poor stars 
in the present sample. 
 At the metal-rich end, a discrepancy persists between a decreasing
[Zn/Fe] with increasing metallicity in the present sample of red giants,
and an approximately constant [Zn/Fe] with metallicity for dwarf bulge
stars.
In conclusion, studies of the 
Galactic bulge with high-resolution spectroscopy for
 several hundred stars such as the present study, as well as work
based on APOGEE data by Schiavon et al. (2017), 
and Schultheis et al. (2017), are crucial to better
understand the chemical evolution and formation of the Galactic bulge.

\begin{acknowledgements}
CRS acknowledges a CAPES/PROEX PhD fellowship.    
  BB and AF acknowledge partial financial support by CNPq, CAPES and FAPESP.
MZ and DM acknowledge
 support by the Ministry of Economy, Development, and Tourism's
Millenium Science Initiative through grant IC120009, awarded to
The Millenium Institute of Astrophysics, MAS, and
from the  BASAL Center for Astrophysics and Associated
 Technologies PFB-06 and FONDECYT Projects 1130196 and 1150345.
SO acknowledges the Italian Ministero dell'Universit\`a e della Ricerca
Scientifica e Tecnologica (MURST), Italy. 
\end{acknowledgements}

\newpage
\begin{appendix}

\section{Comparison between GIRAFFE and UVES spectra}

Figures \ref{zn1} to \ref{zn12} shown in appendices A.1 to A.12
present the  fits of the \ion{Zn}{I} 6362.3 {\rm \AA} line,
for both spectra GIRAFFE and UVES for stars in common between the two
sets of observations.

\begin{figure}
\centering
\includegraphics[width=\columnwidth]{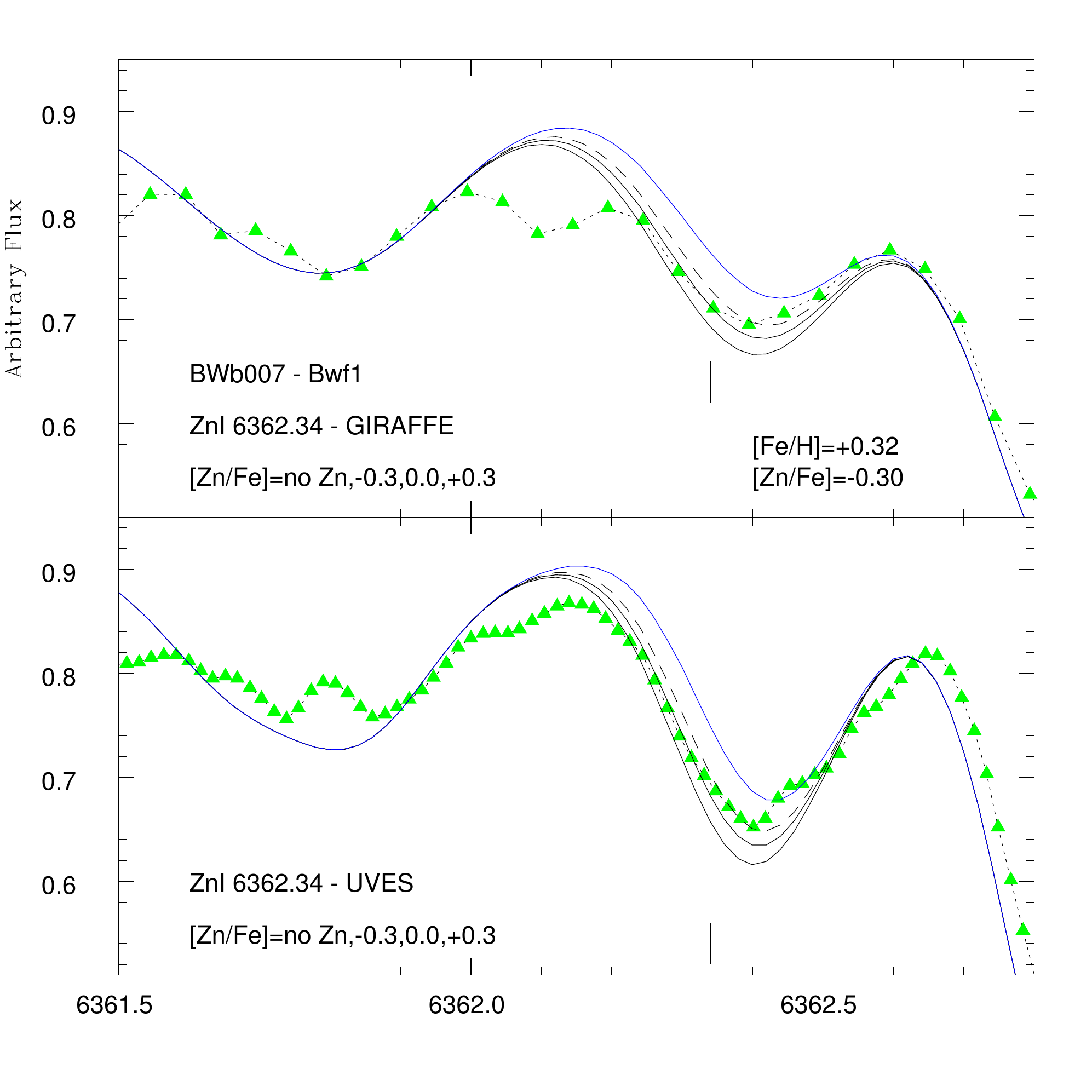}
\includegraphics[width=\columnwidth]{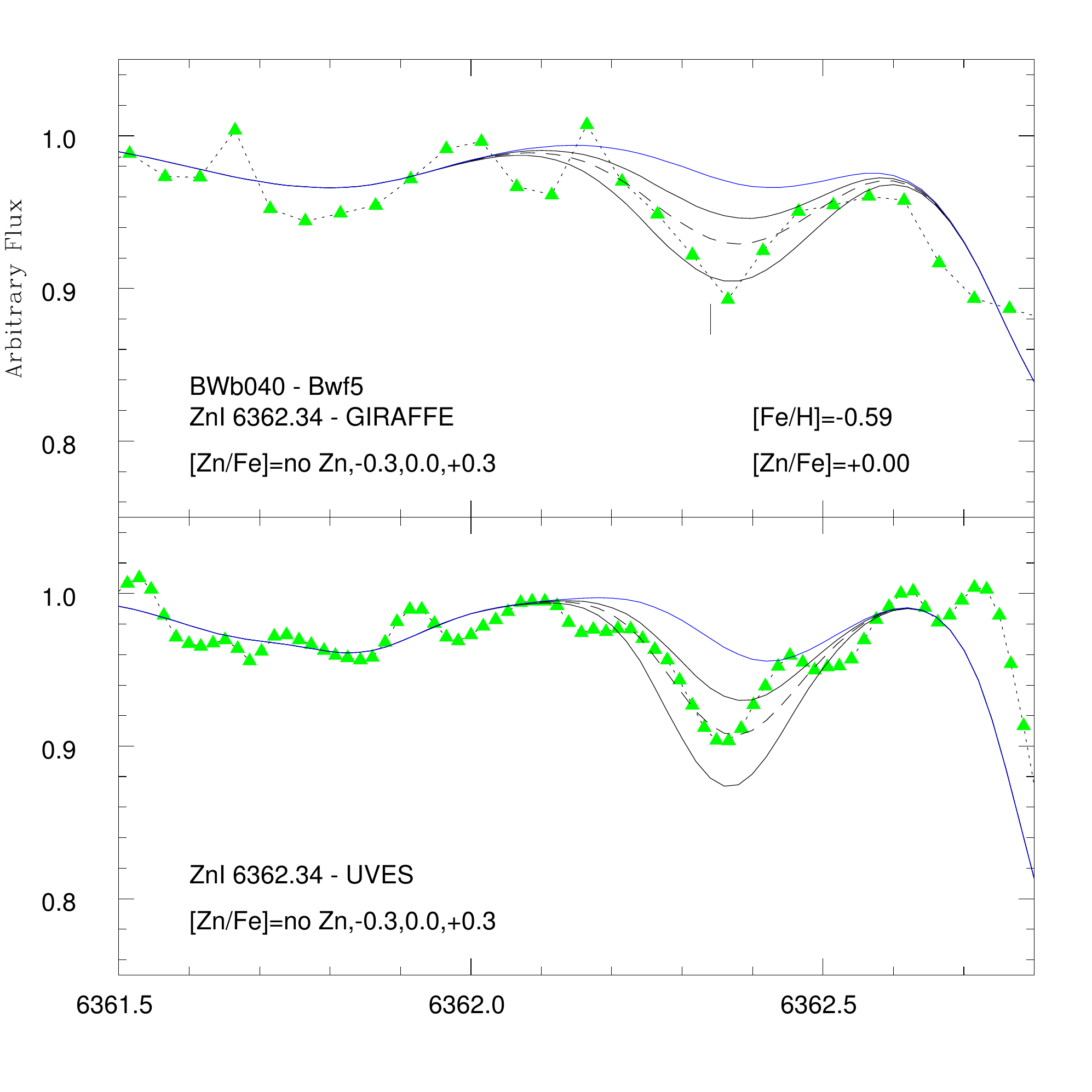}
\caption{Comparison between UVES and Giraffe spectra with
fits of the \ion{Zn}{I} 6362.3 {\rm \AA} line, for
stars in common. Symbols: dotted black line and green filled triangles
correspond to the observed spectra: black lines: synthetic spectra,
dashed line: synthetic spectrum for the chosen [Zn/Fe] value.
Blue solid line: synthetic spectra without Zn, showing the CN line. }
\label{zn1} 
\end{figure}

\begin{figure}
\centering
\includegraphics[width=\columnwidth]{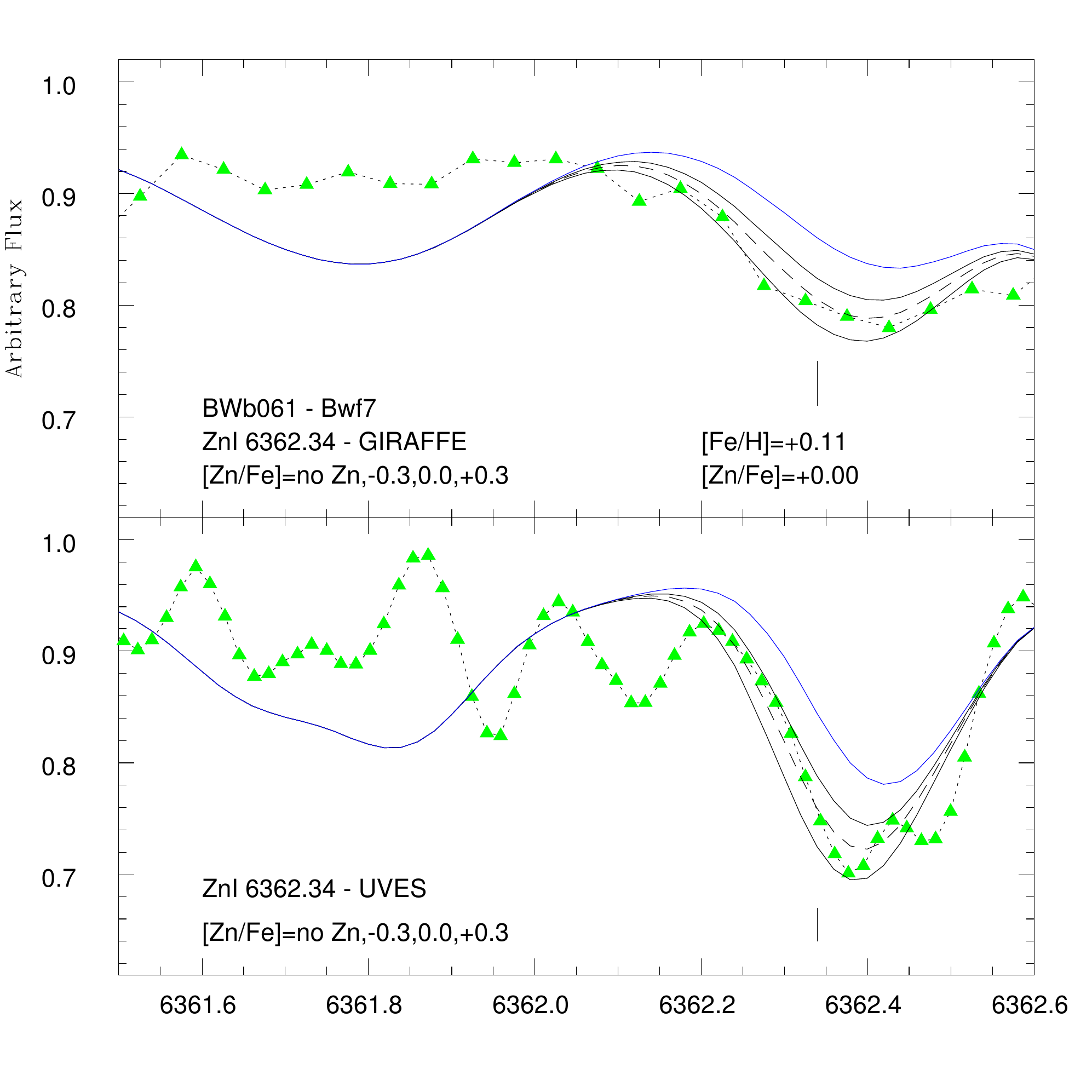}
\includegraphics[width=\columnwidth]{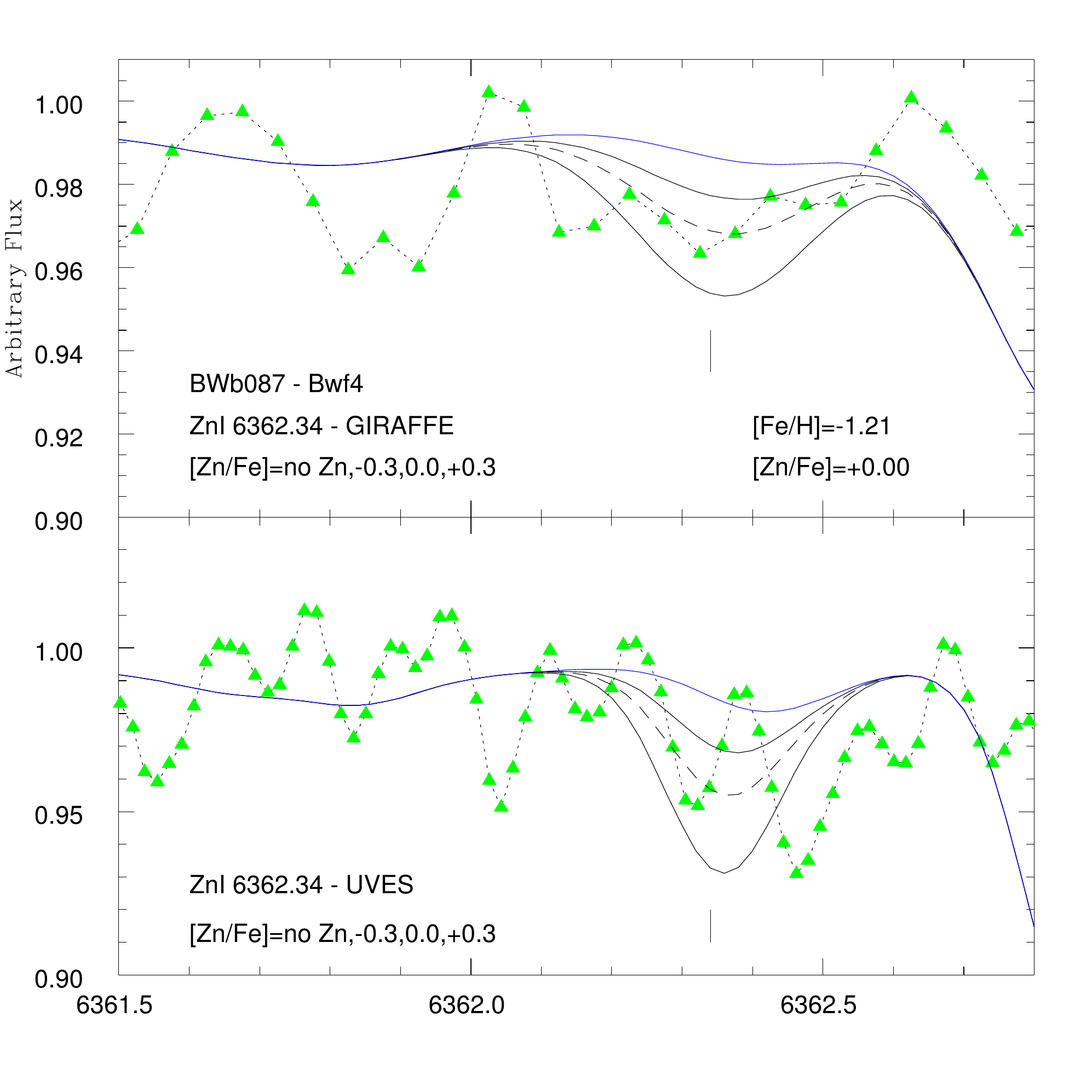}
\caption{Continuation of Fig. \ref{zn1} }
\label{zn2} 
\end{figure}

\begin{figure}
\centering
\includegraphics[width=\columnwidth]{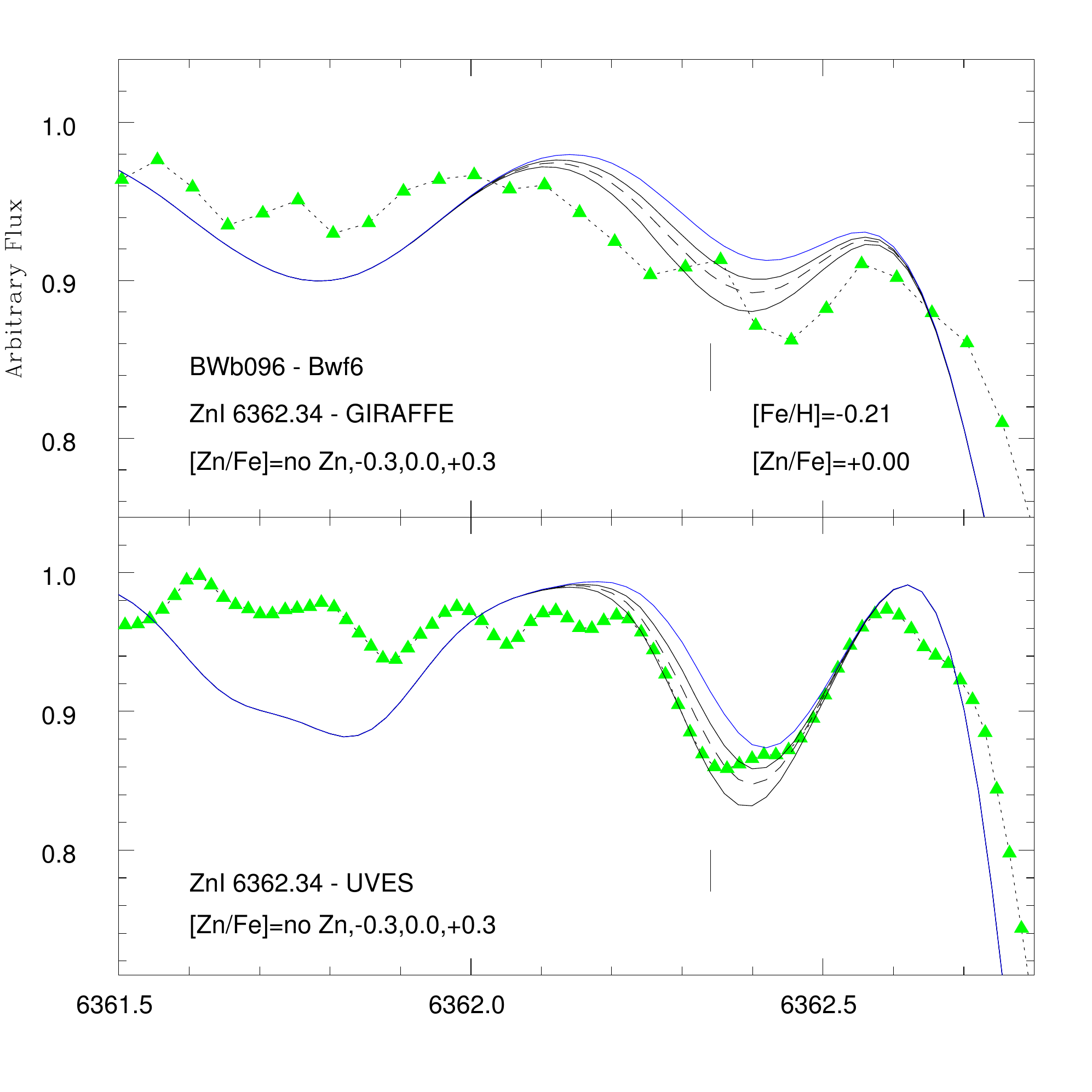}
\includegraphics[width=\columnwidth]{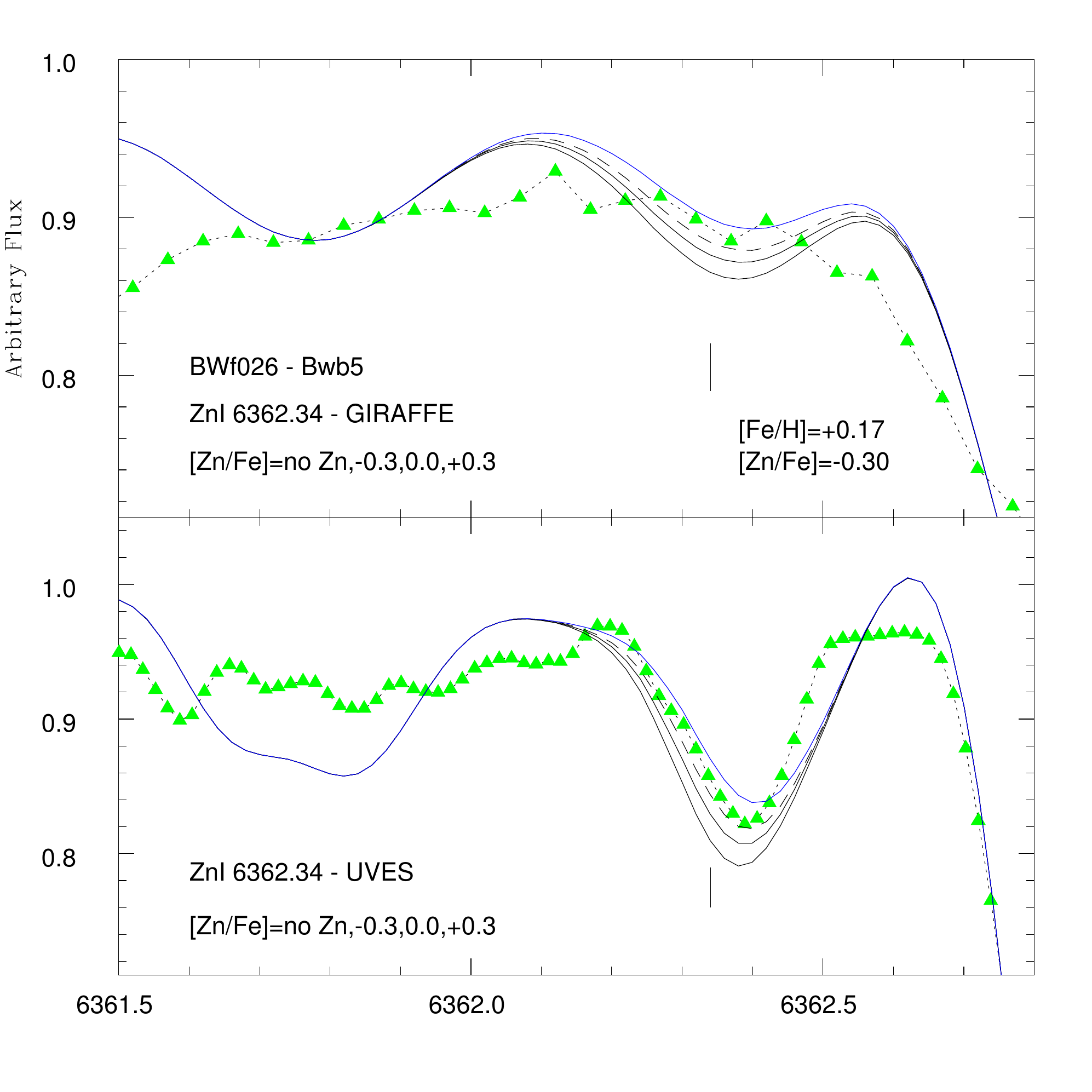}
\caption{Continuation of Fig. \ref{zn1} }
\label{zn3} 
\end{figure}

\begin{figure}
\centering
\includegraphics[width=\columnwidth]{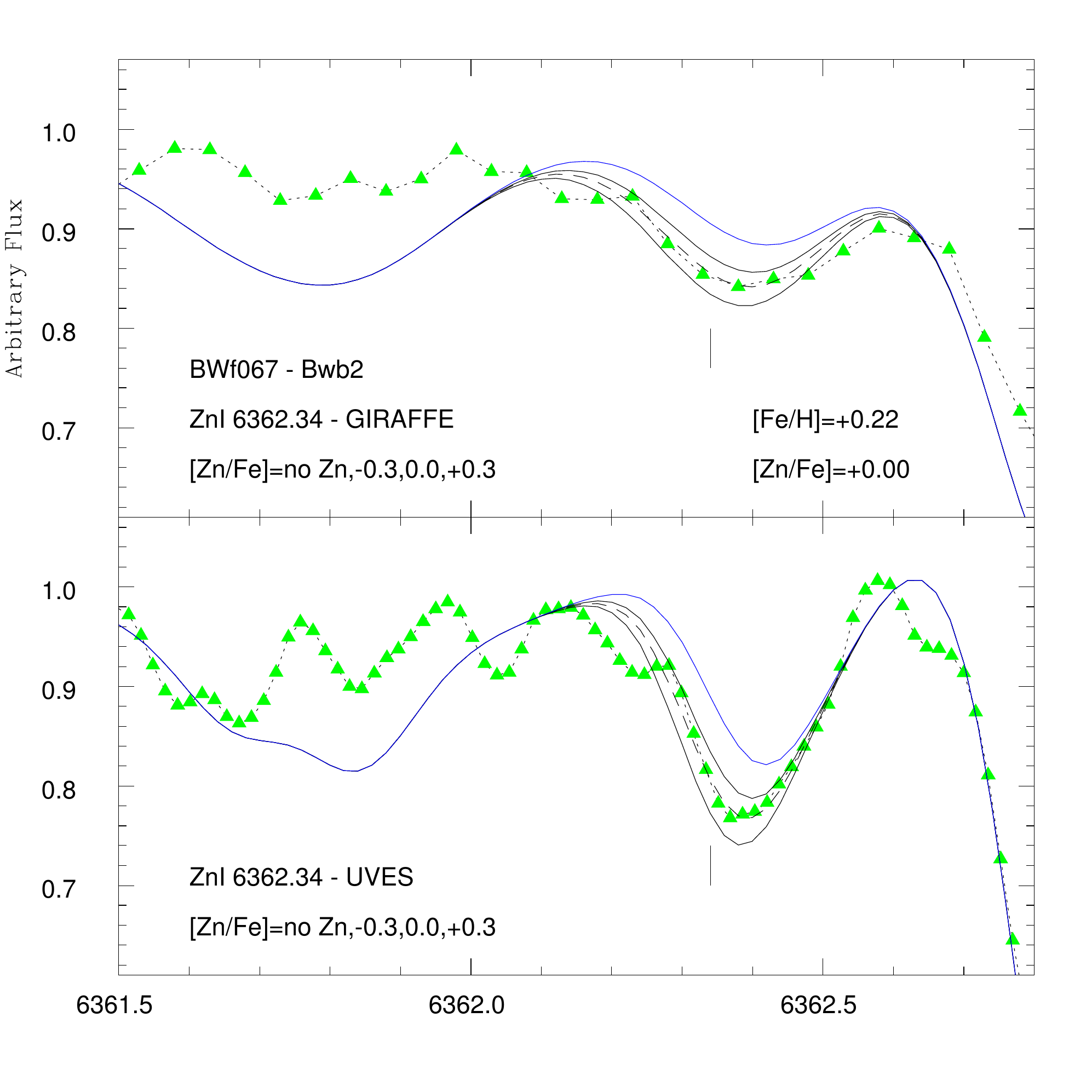}
\includegraphics[width=\columnwidth]{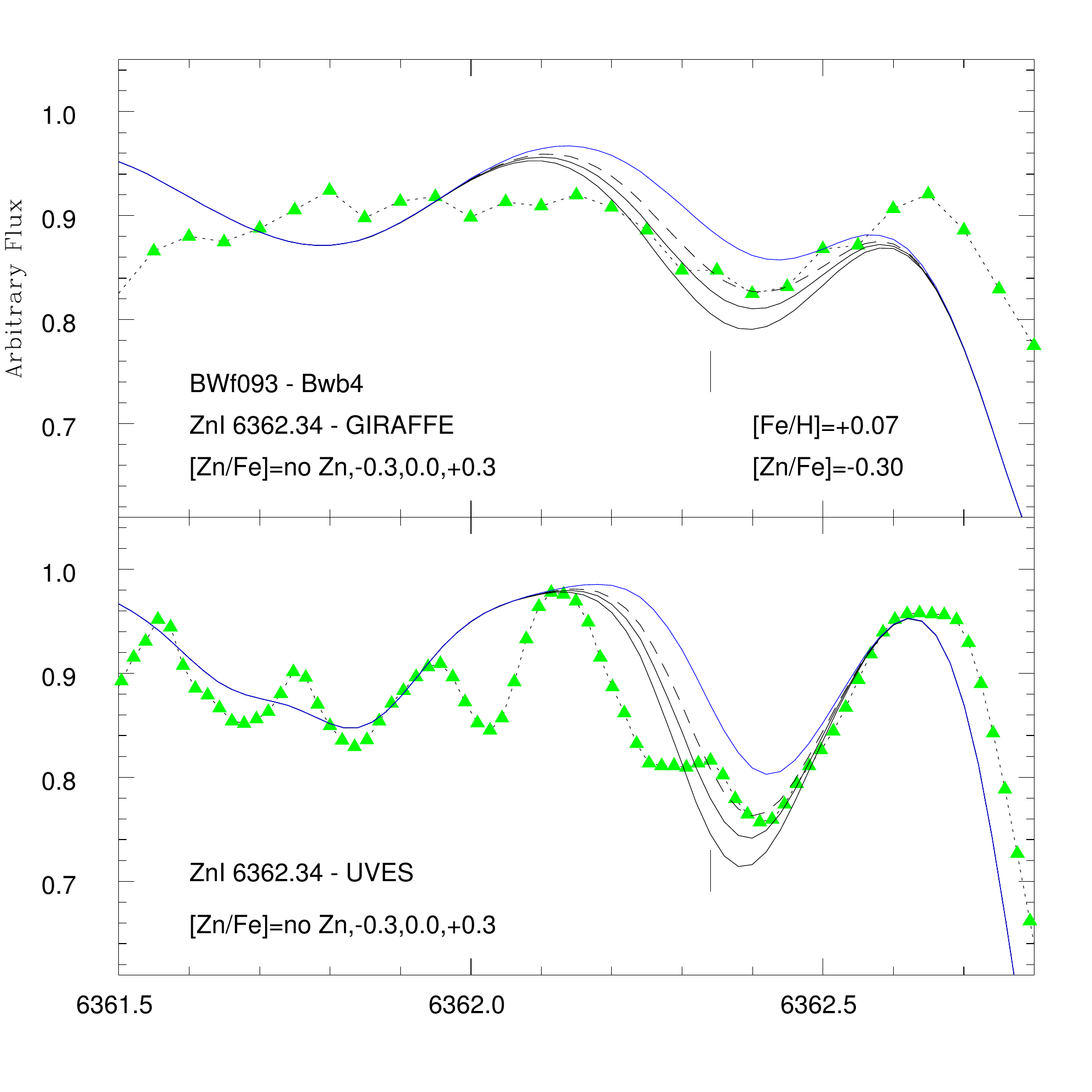}
\caption{Continuation of Fig. \ref{zn1} }
\label{zn4} 
\end{figure}

\begin{figure}
\centering
\includegraphics[width=\columnwidth]{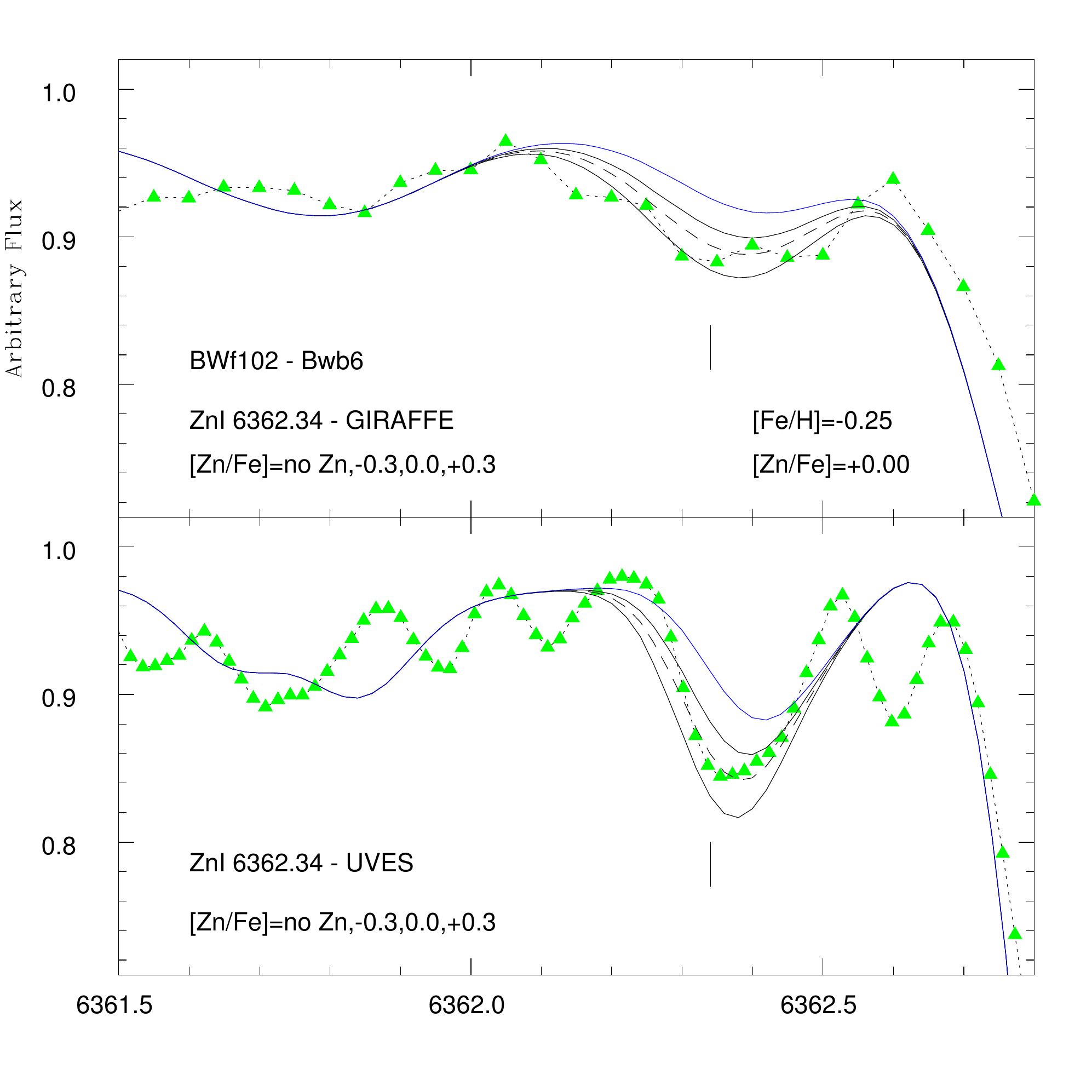}
\includegraphics[width=\columnwidth]{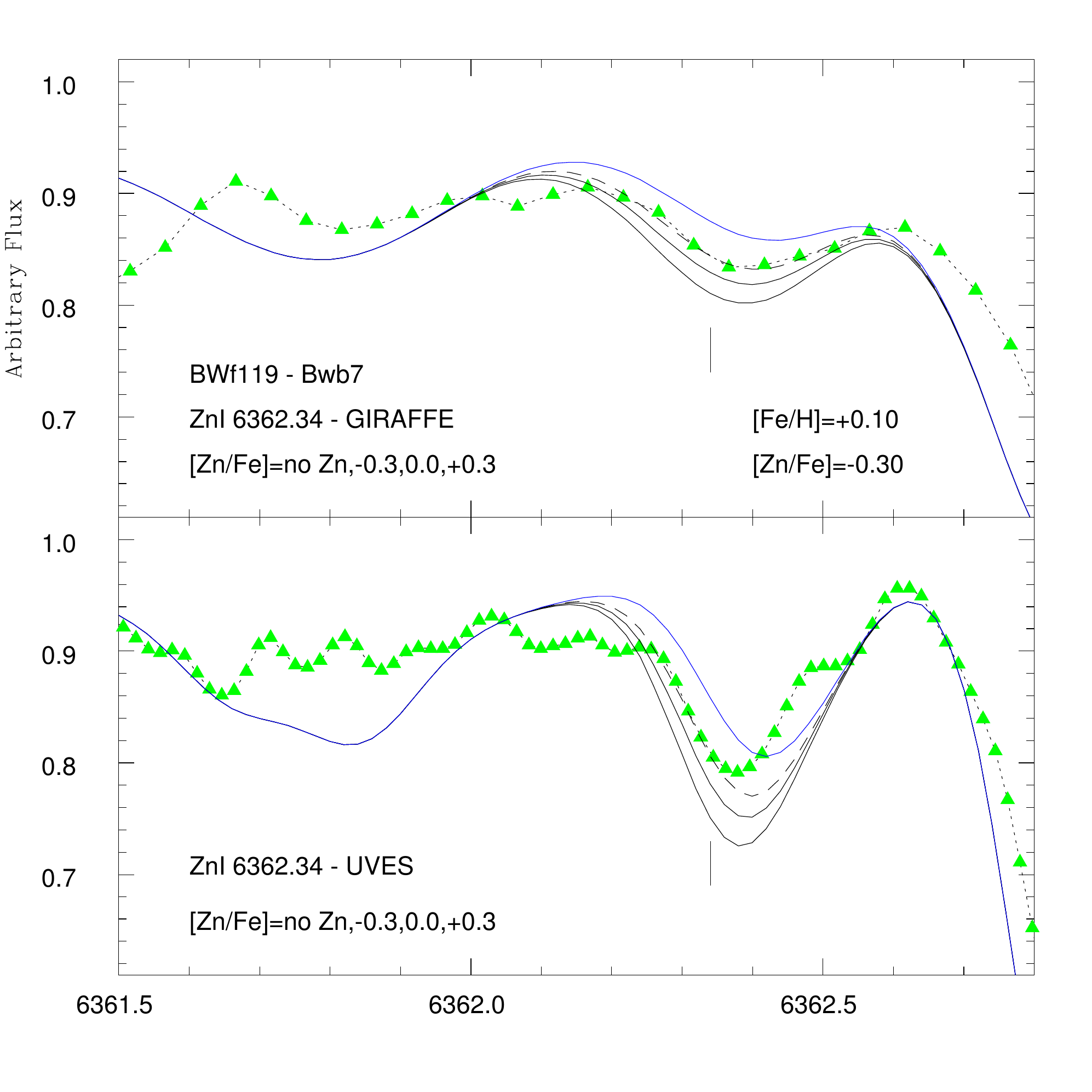}
\caption{Continuation of Fig. \ref{zn1} }
\label{zn5} 
\end{figure}

\begin{figure}
\centering
\includegraphics[width=\columnwidth]{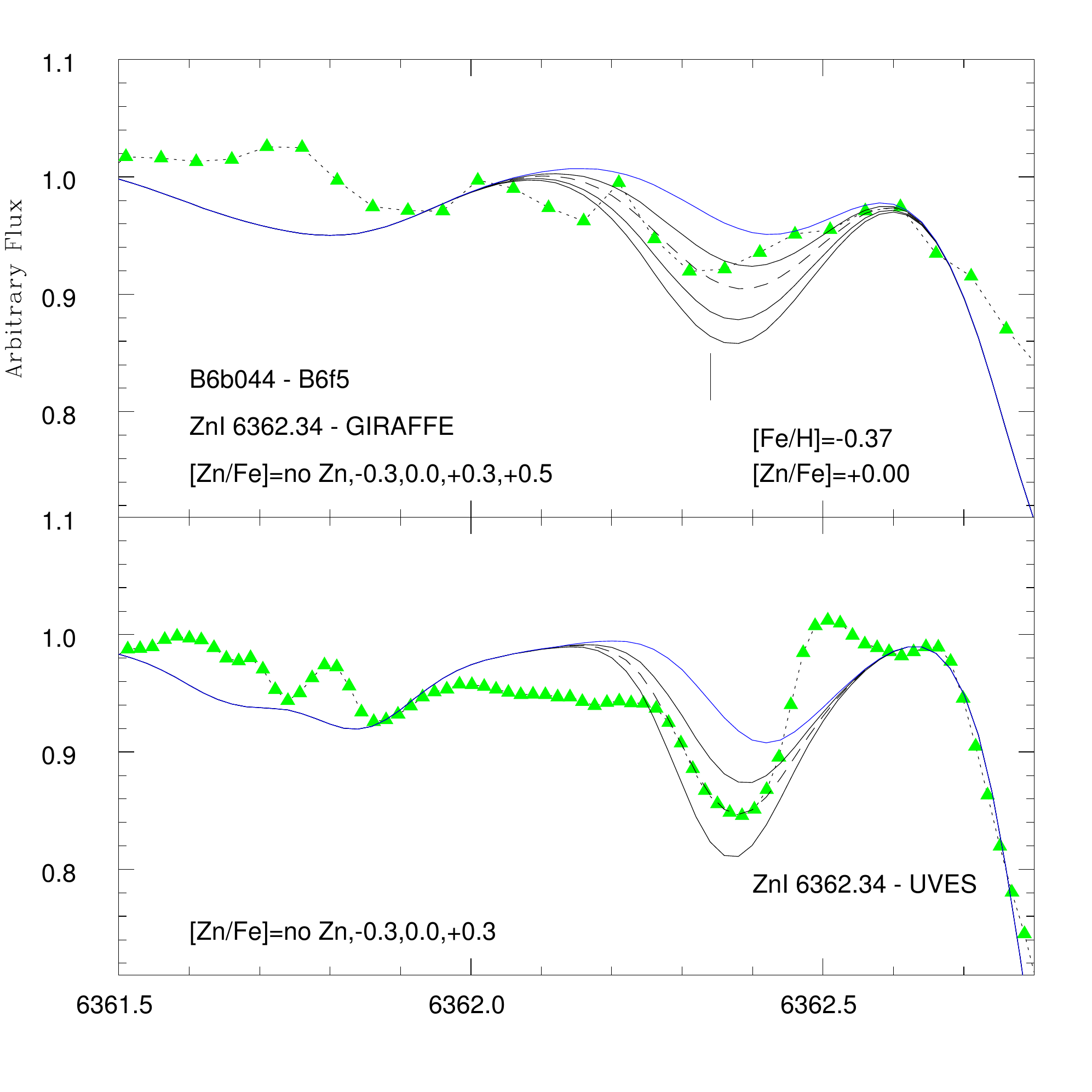}
\includegraphics[width=\columnwidth]{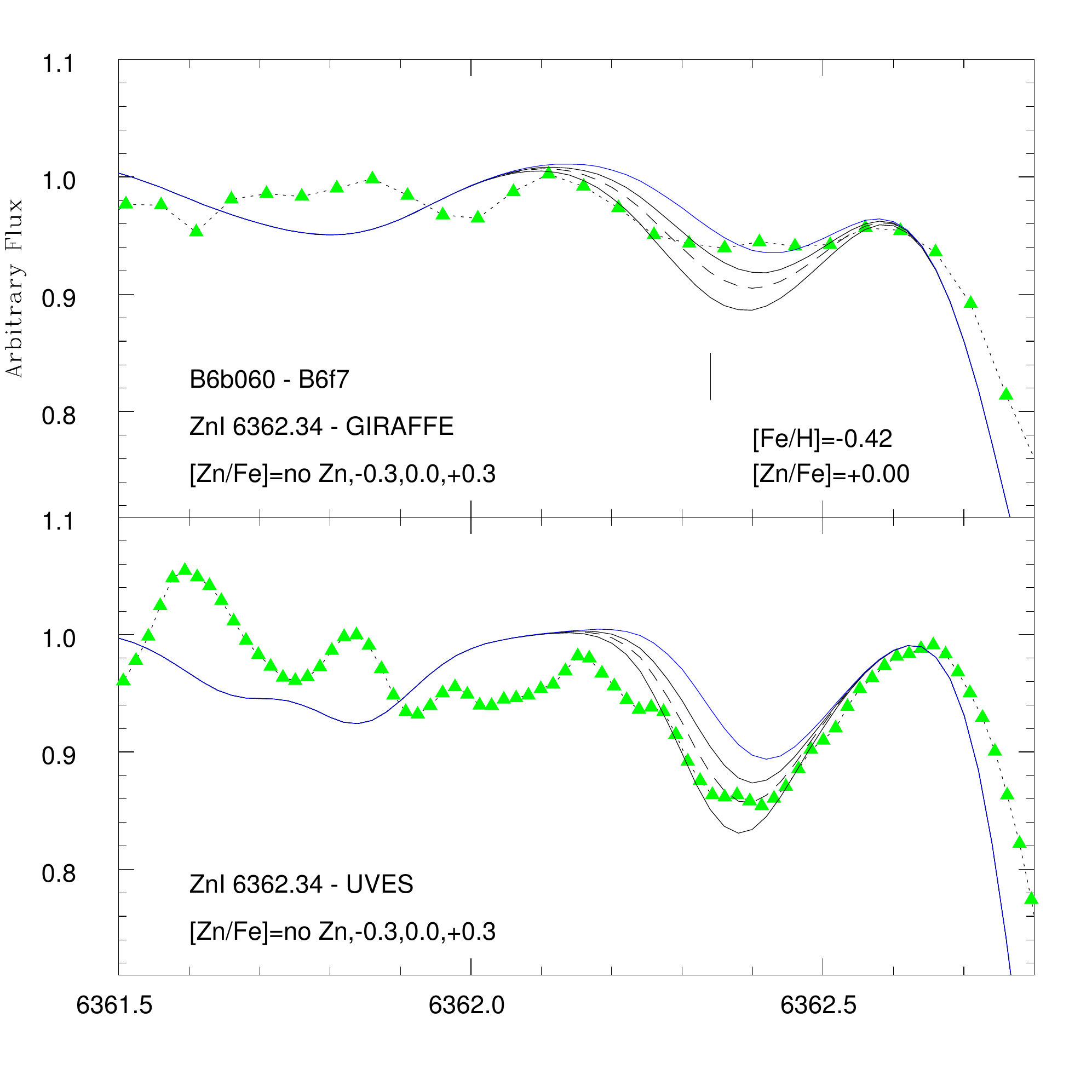}
\caption{Continuation of Fig. \ref{zn1} }
\label{zn6} 
\end{figure}

\begin{figure}
\centering
\includegraphics[width=\columnwidth]{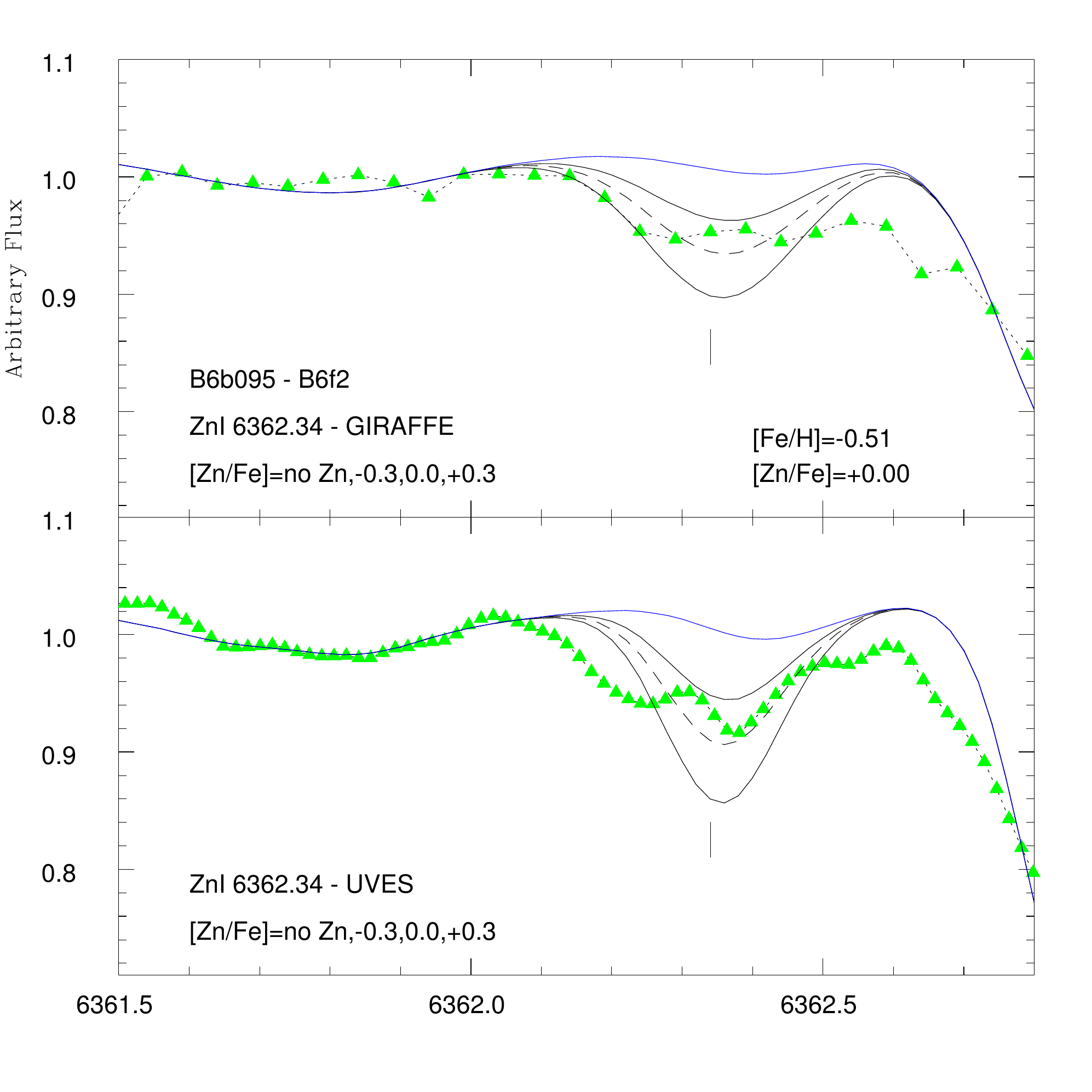}
\includegraphics[width=\columnwidth]{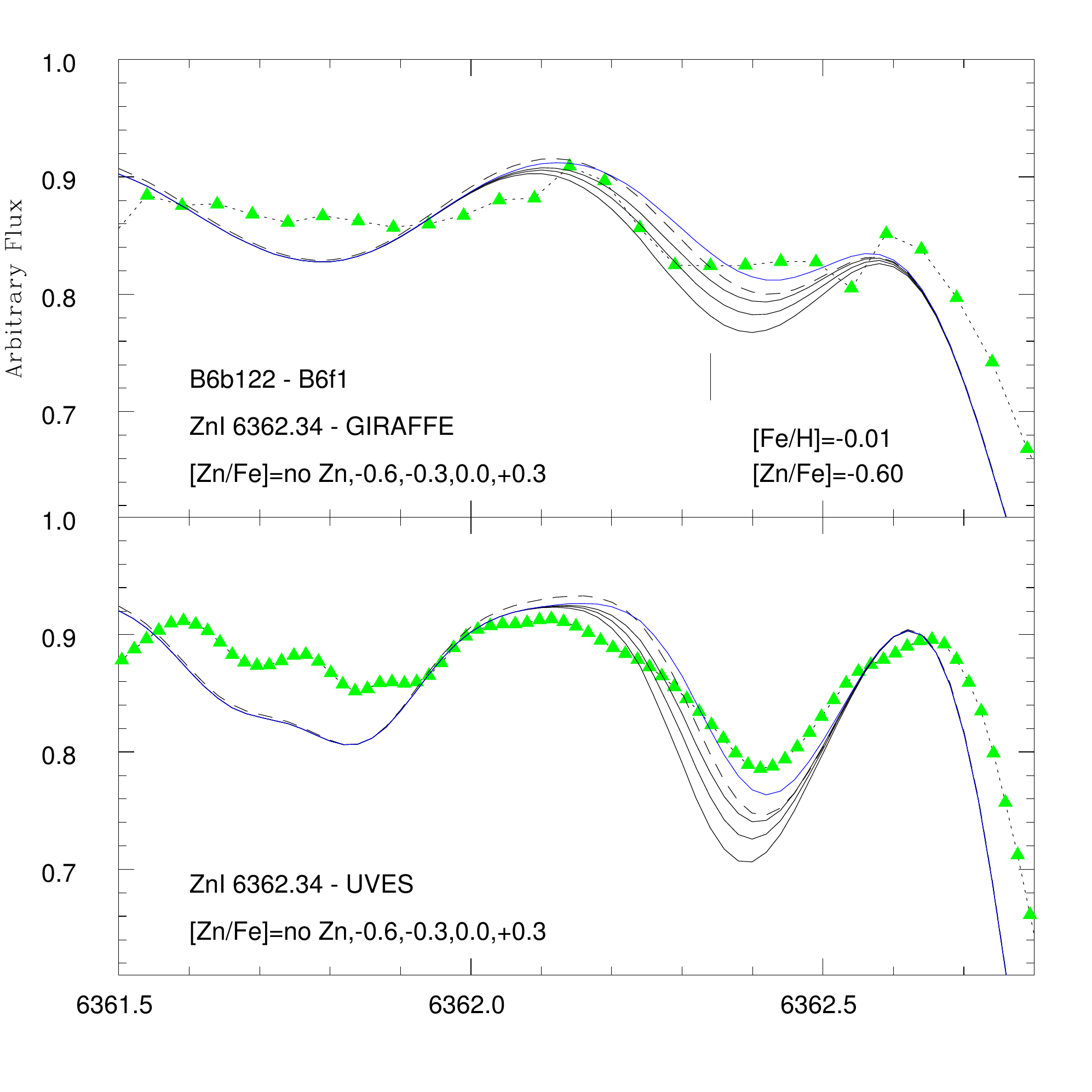}
\caption{Continuation of Fig. \ref{zn1} }
\label{zn7} 
\end{figure}

\begin{figure}
\centering
\includegraphics[width=\columnwidth]{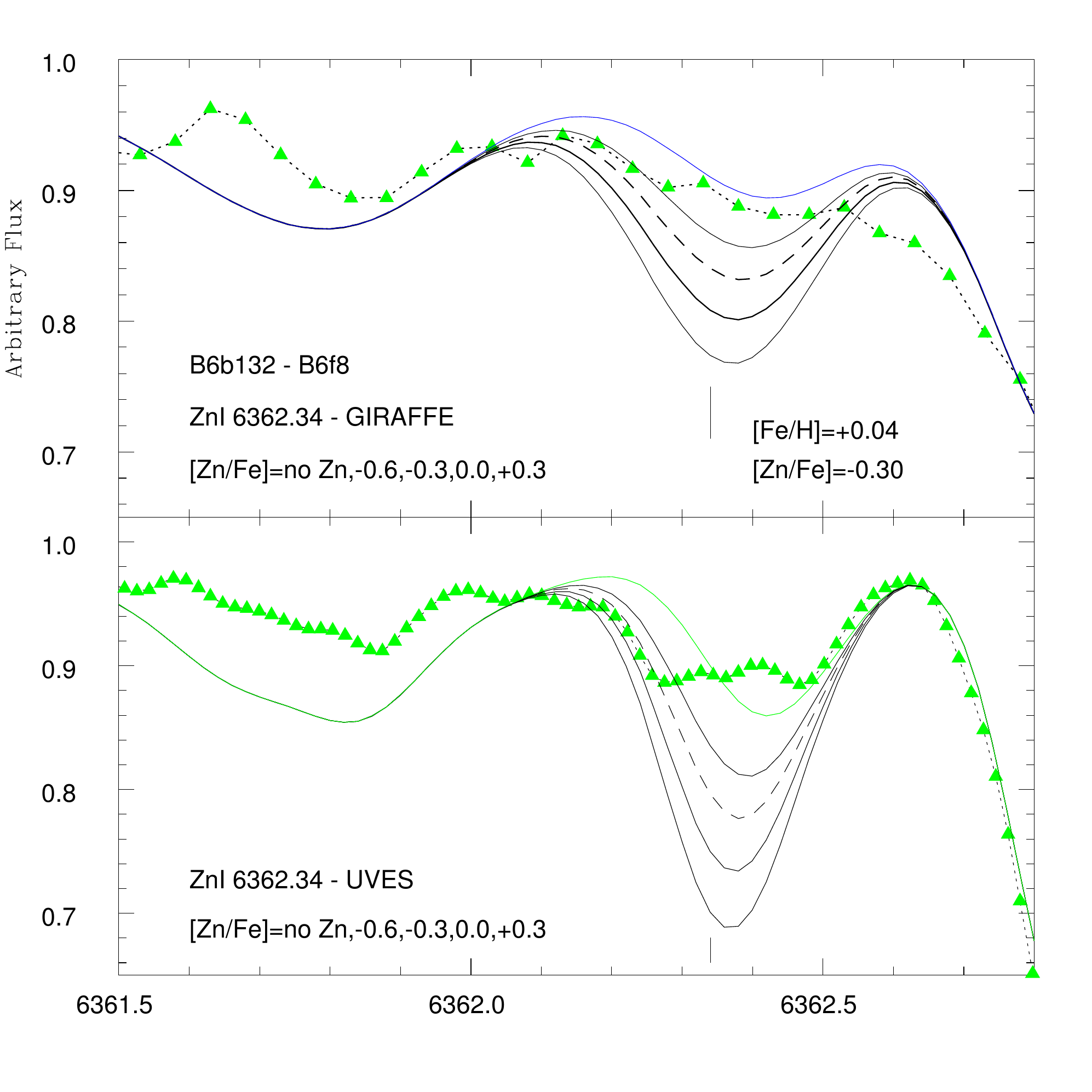}
\includegraphics[width=\columnwidth]{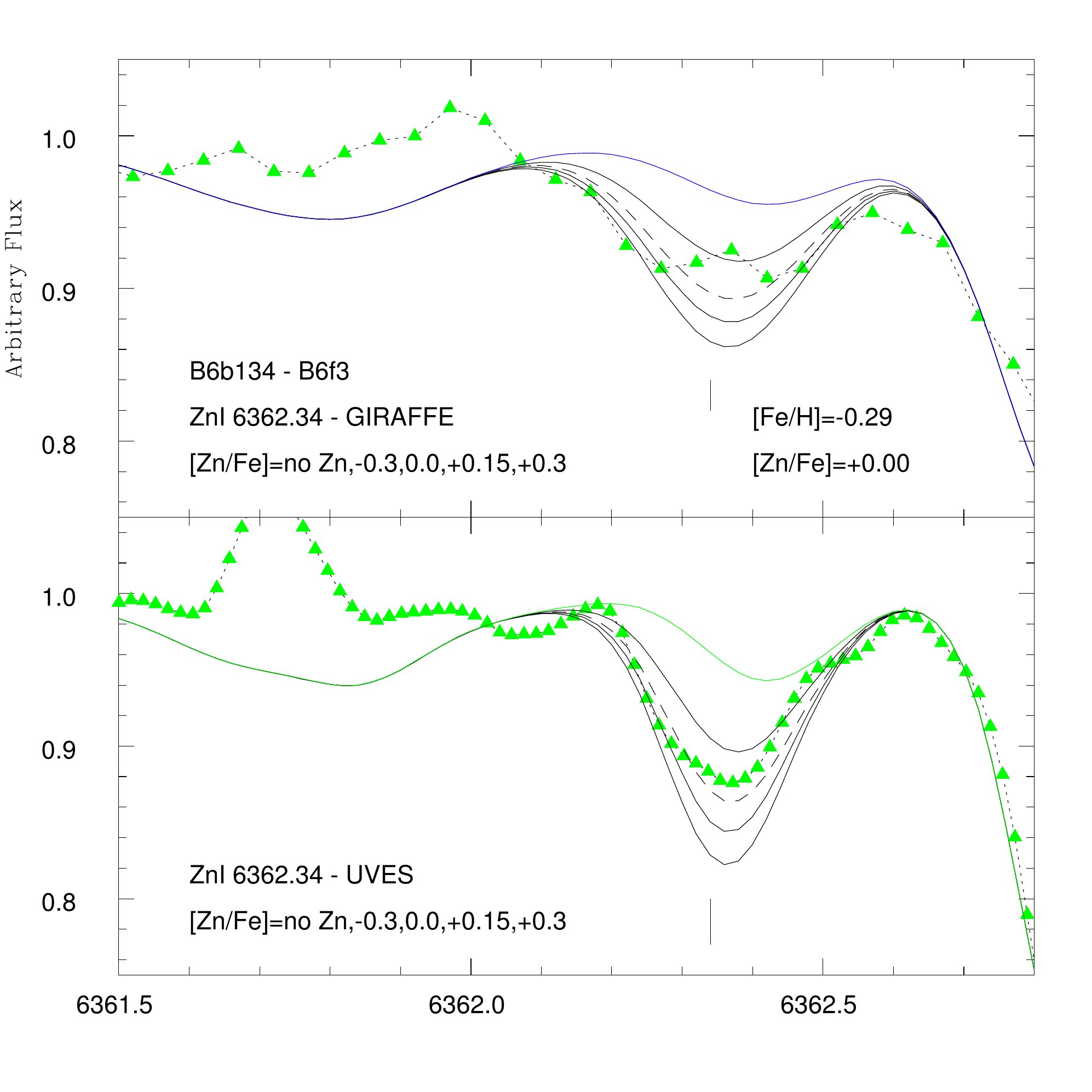}
\caption{Continuation of Fig. \ref{zn1} }
\label{zn8} 
\end{figure}

\begin{figure}
\centering
\includegraphics[width=\columnwidth]{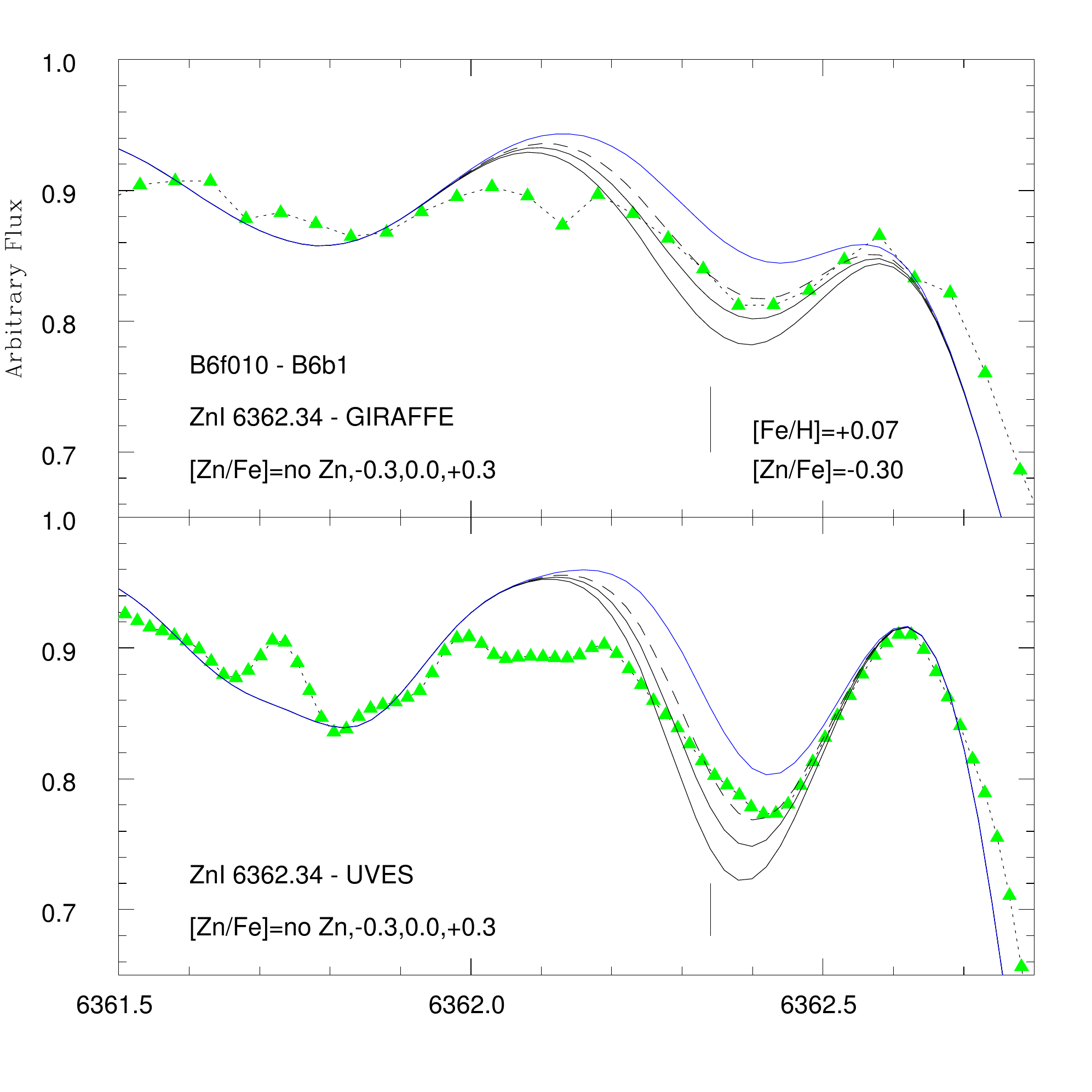}
\includegraphics[width=\columnwidth]{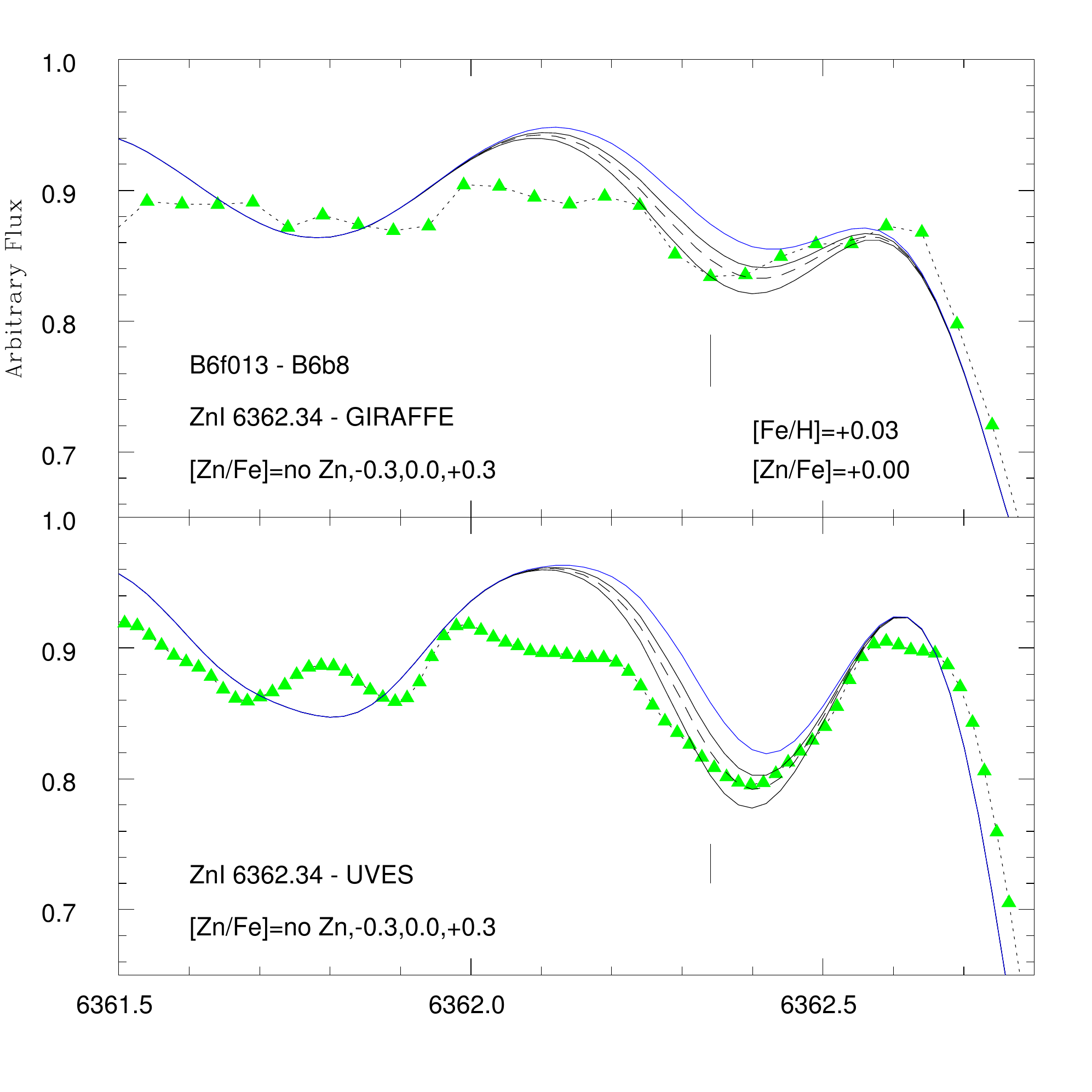}
\caption{Continuation of Fig. \ref{zn1} }
\label{zn9} 
\end{figure}

\begin{figure}
\centering
\includegraphics[width=\columnwidth]{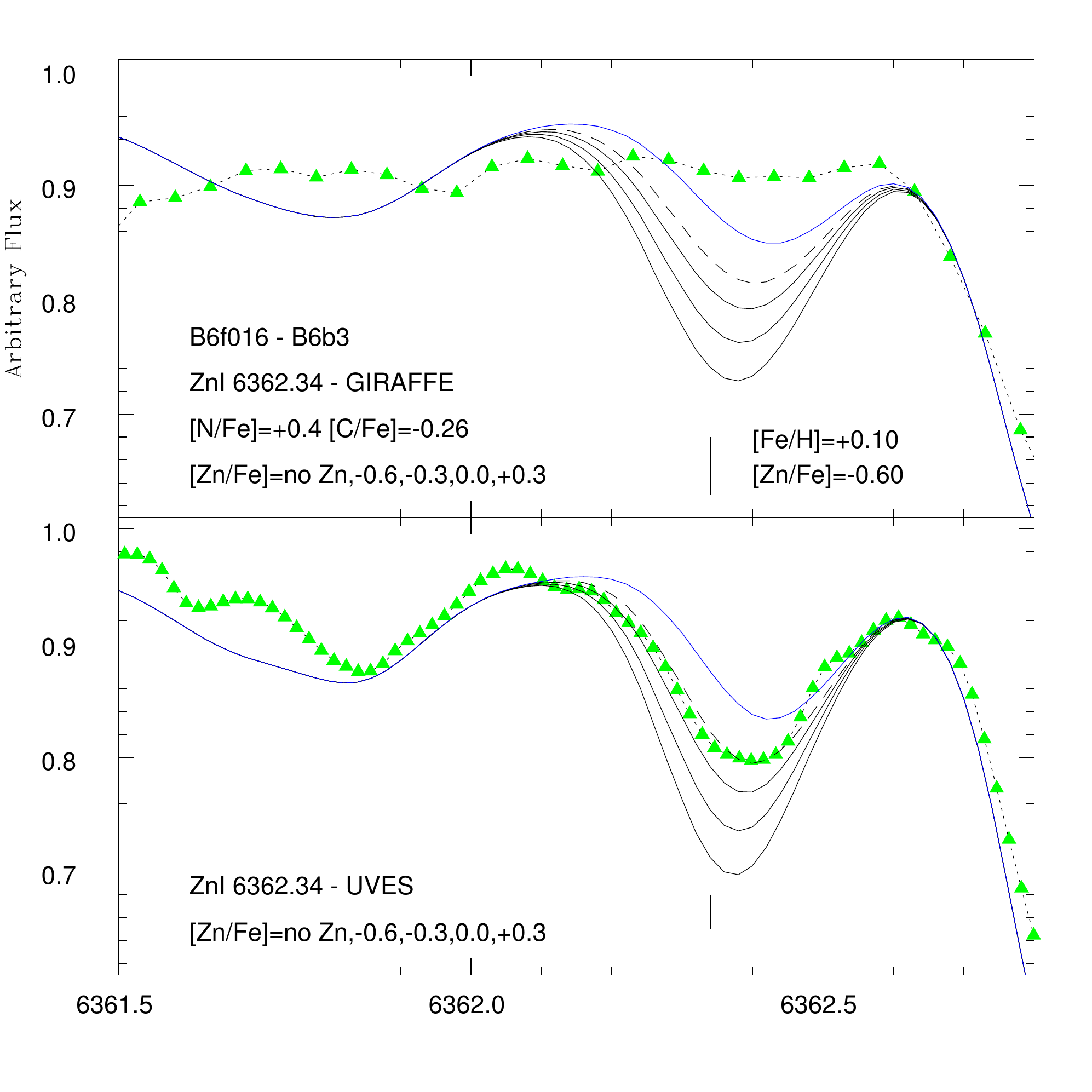}
\includegraphics[width=\columnwidth, keepaspectratio]{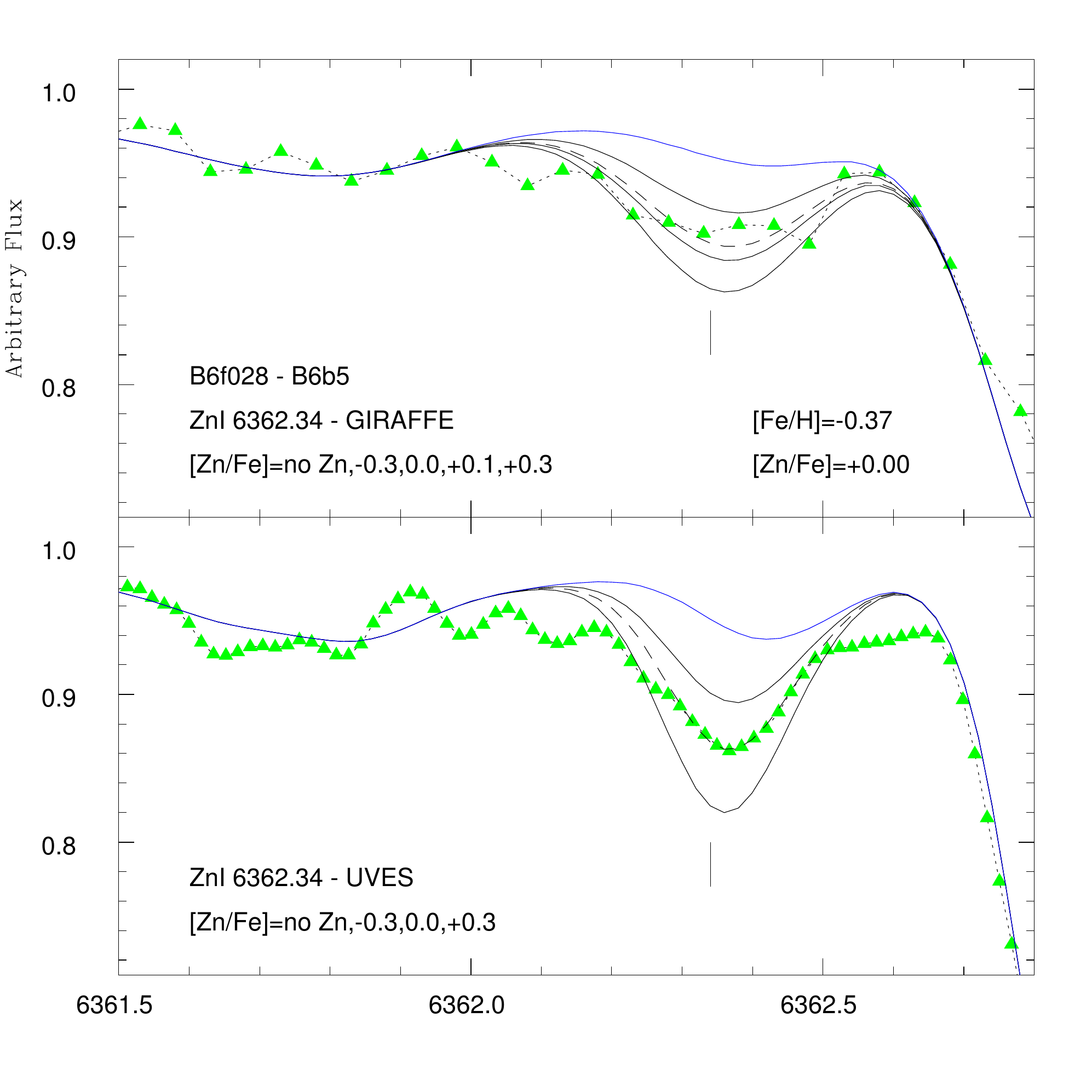}
\caption{Continuation of Fig. \ref{zn1} }
\label{zn10} 
\end{figure}

\begin{figure}
\centering
\includegraphics[width=\columnwidth]{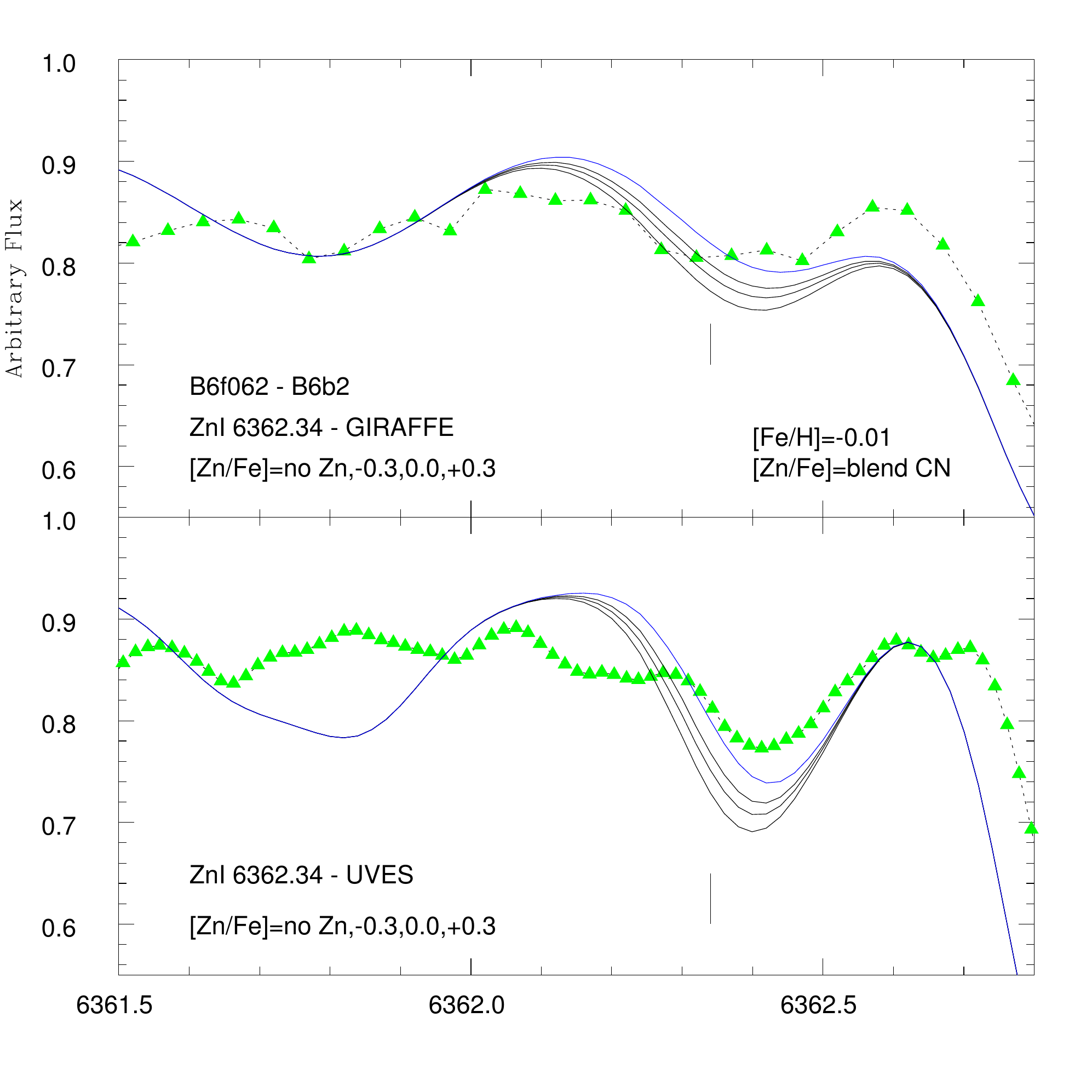}
\includegraphics[width=\columnwidth]{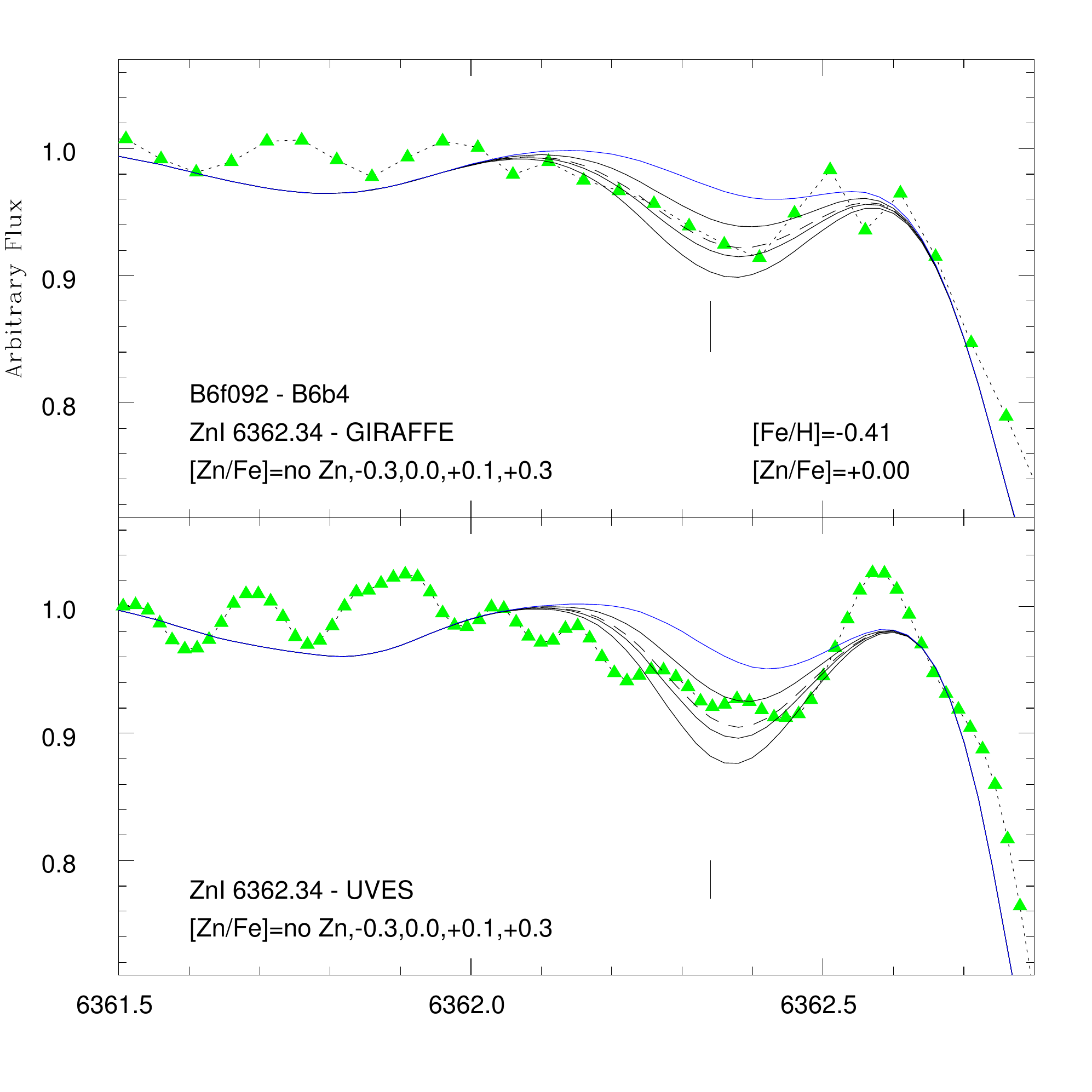}
\caption{Continuation of Fig. \ref{zn1} }
\label{zn11} 
\end{figure}

\begin{figure}
\centering
\includegraphics[width=\columnwidth]{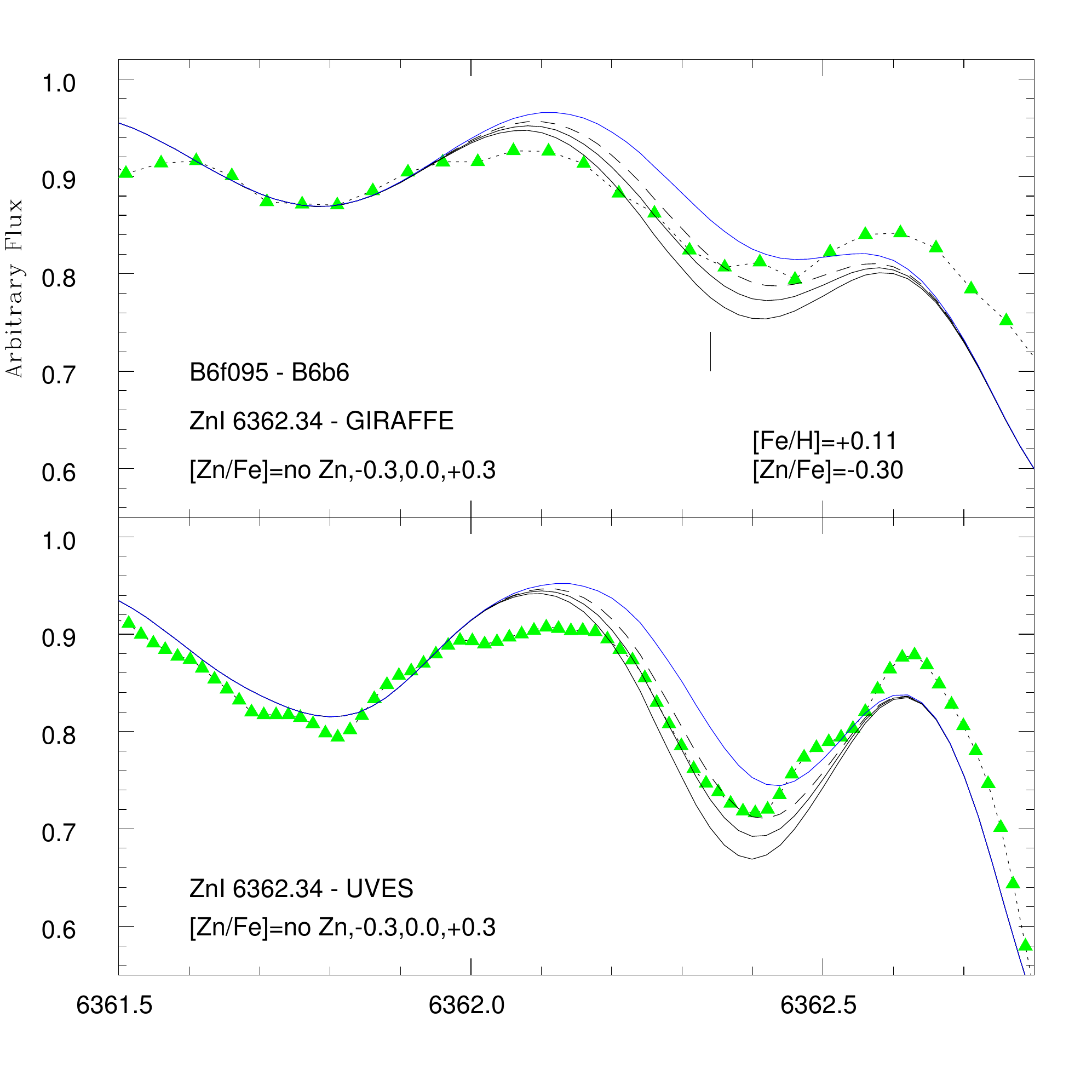}
\caption{Continuation of Fig. \ref{zn1} }
\label{zn12} 
\end{figure}

\section{Final abundances.}

\scalefont{0.6}
\begin{longtable}{llrrrrrrrrrrrrr}
\caption{\label{results}OGLE and GIRAFFE names, stellar parameters, 
resulting [N/Fe], [O/Fe], [Zn/Fe],
and alpha-element abundances. For stars having both GIRAFFE and UVES
spectra, the parameters from both UVES and GIRAFFE analyses are reported.
These stars are marked with an asterisk when corresponding to the UVES results.} \\
\hline
\endfirsthead
\caption{continued.}\\
\hline\hline
\hbox{OGLE n$^{\circ}$} & \hbox{GIRAFFE n$^{\circ}$} &  \hbox{T$_{\rm eff}$} & \hbox{log~g} 
& \hbox{[Fe/H]} & \hbox{v$_{\rm t}$} & \hbox{[N/Fe]} & \hbox{[O/Fe]} 
& \hbox{[Zn/Fe]} & \hbox{[Mg/Fe]} & \hbox{[Si/Fe]} & \hbox{[Ca/Fe]}
& \hbox{[Ti/Fe]} & \\
\hline
\endhead
\hline
\endfoot
\noalign {}
\hline
\noalign {}
OGLE    &       GIRAFFE &\hbox{T$_{eff}$}&\hbox{log~g}  & \hbox{[Fe/H]} &\hbox{v$_{\rm t}$}    &\hbox{[N/Fe]}  &\hbox{[O/Fe]}  &\hbox{[Zn/Fe]} &\hbox{[Mg/Fe]} &\hbox{[Si/Fe]} &\hbox{[Ca/Fe]} &\hbox{[Ti/Fe]} & \\
\noalign {}
\hline
\noalign {\vskip 0.05cm}
\noalign {\hskip 2.0cm Baade's window bright: BW-f}
\noalign {\vskip 0.05cm}
\hline
\noalign {}
423342  &       bwb002  &       4650    &       1.99    &       0.46    &       1.3     &       -0.05   &       -0.35   &       -0.30   &       -0.04   &       -0.08   &       0.13    &       0.03    & \\
423323  &       bwb003  &       4200    &       1.59    &       -0.48   &       1.5     &       0.00    &       0.10    &       0.22    &       0.43    &       0.26    &       0.15    &       0.23    & \\
412779  &       bwb004  &       4850    &       1.93    &       -0.37   &       1.5     &       0.20    &       ----    &       0.13    &       0.23    &       0.22    &       0.29    &       0.48    & \\
412803  &       bwb005  &       4000    &       1.52    &       0.51    &       1.3     &       -0.20   &       -0.35   &       -0.40   &       ----    &       ----    &      ----       &      ----     & \\
423359  &       bwb006  &       4650    &       1.92    &       -1.23   &       1.4     &       ----    &       ----    &       0.30    &       0.34    &       0.43    &       0.30    &       0.48    & \\
433669* &       bwb007* &       4400    &       1.80    &       0.32    &       1.6     &       0.45    &       -0.18   &       -0.30   &       -0.02   &       -0.10   &       0.01    &       -0.04   & \\
433669  &       bwb007  &       4300    &       1.67    &       0.32    &       1.5     &       0.30    &       -0.25   &       -0.40   &       -0.02   &       -0.10   &       0.01    &       -0.04   & \\
\end{longtable}

\end{appendix}


\begin{thebibliography}{}
\bibitem[]{} Akerman, C.J., Ellison, S.L., Pettini, M., Steidel, C.C. 2005,
A\&A, 440, 499
\bibitem[]{alonso99} Alonso, A., Arribas, S., Mart\'{\i}nez-Roger, C.
        1999, A\&AS, 140, 261 
\bibitem[]{} Alves-Brito, A., Mel\'endez, J., Asplund, M. et al. 2010,
    A\&A, 513, A35
\bibitem[Asplund et al.(2009)]{asplund09}  Asplund, M., Grevesse, N., Sauval, A.J., Scott, P. 2009,
ARA\&A, 47, 481
\bibitem[]{} Barbuy, B. 1988, A\&A, 191, 121
\bibitem[]{} Barbuy, B., Perrin, M.-N., Katz, D., Coelho, P., Cayrel, R.,
Spite, M., van't Veer-Menneret, C. 2003, A\&A, 404, 661 
\bibitem[]{} Barbuy, B., Hill, V., Zoccali, M. et al. 2013, A\&A, 559, A5
\bibitem[]{} Barbuy, B., Fria\c ca, A., da Silveira, C.R. et al. 2015, 
A\&A, 580, A40
\bibitem[]{} Barbuy, B., Chiappini, C., Gerhard, O. 2018, ARA\&A, in press

\bibitem[]{} Bensby, T., Feltzing, S., Lundstr\"om, I. 2004, A\&A, 415, 155
\bibitem[]{} Bensby, T., Yee, J.C., Feltzing, S. et al. 2013, A\&A, 549, A147
\bibitem[]{} Bensby, T., Feltzing, S., Oey, M.S. 2014, A\&A, 562, A71
\bibitem[]{} Bensby, T., Feltzing, S., Gould, A. et al. 2017, A\&A, 605, A89
\bibitem[]{} Carpenter, J.M. 2001, AJ, 121, 2851
\bibitem[]{} Carretta, E., Bragaglia, A., Gratton, R.G.
et al. 2009, A\&A, 505, 117 
\bibitem[]{} Casey, A.R, Schlaufman, K.C. 2015, ApJ, 809, 110
\bibitem[]{} Cavichia, O., Moll\'a, M., Costa, R.D.D., Maciel, W.J. 
  2014, MNRAS, 437, 3688

\bibitem[]{} Cayrel, R., Depagne, E., Spite, M. et al. 2004, A\&A, 416, 1117
\bibitem[]{}
Cescutti G, Matteucci F, Lanfranchi GA, McWilliam A. 2008, A\&A, 491, 401
\bibitem[]{}
Cescutti G, Chiappini C, Hirshi R, et~al. 2018, in prep.

\bibitem[]{} Coelho, P., Barbuy, B., Mel\'endez, J., Schiavon, R.P.,
 Castilho, B.V.  2005, A\&A, 443, 735
\bibitem[]{} Cooke, R., Pettini, M., Jorgenson, R.A., Murphy, M.T.,
   Rudie, G.C., Steidel, C.C. 2013, MNRAS, 431, 1625
\bibitem[]{} Cooke, R.J., Pettini, M., Jorgenson, R.A. 2015, ApJ, 800, 12
\bibitem[]{} Cunha, K., Smith, V.V. 2006, ApJ, 652, 491
\bibitem[]{} Davis, S.P., Phillips, J.G. 1963, The Red System 
(A$^{2}\Pi$-X$^{2}\Sigma$) of the CN molecule

\bibitem[]{}
Duffau, S., Caffau, E., Sbordone, L. et al. 2017, A\&A, 605, A128

\bibitem[Fenner et al. (2004)]{fenner04}
Fenner, Y., Prochaska, J.X., Gibson, B.K 2004, ApJ, 606, 116
\bibitem[Fria\c ca \& Barbuy (2017)]{friacabarbuy17}
Fria\c ca, A.C.S., Barbuy, B. 2017, A\&A, 598, A121 (FB17)

\bibitem[]{} Fulbright, J.P., McWilliam, A., Rich, R.M. 2007, ApJ, 661, 1152 
\bibitem[]{} Gaia collaboration, 2017, A\&A, 605, A79
\bibitem[Garcia-Perez et al. (2013)]{garcia-perez13}
Garc\'{\i}a-P\'erez, A.E., Cunha, K., Shetrone, M. et al. 2013, ApJ, 767, L9
\bibitem[]{} Gonzalez, O.A., Rejkuba, M., Zoccali, M. et al. 2011, A\&A,
530, A54
\bibitem[]{} Gustafsson, B., Edvardsson, B., Eriksson, K. et al. 2008, A\&A, 486, 951

\bibitem[]{} Hawkins, K., Jofr\'e, P., Masseron, T., Gilmore, G. 2015, MNRAS, 453, 758
\bibitem[]{} Hill, V., Lecureur, A., G\'omez, A. et al. 2011,
A\&A, 534, A80

\bibitem[Howes et al. (2014)]{howes14}
Howes, L.M., Asplund, M., Casey, A.R. et al. 2014, MNRAS, 445, 4241

\bibitem[Howes (2015)]{howestese15}
Howes, L.M. 2015a, PhD thesis, Australian National University

\bibitem[Howes et al. (2015)]{howes15}
Howes, L.M., Casey, A.R., Asplund, M. et al. 2015b, Nature, 527, 484

\bibitem[Howes et al. (2016)]{howes16}
Howes, L.M., Asplund, M., Keller, S.C. et al. 2016, MNRAS, 460, 884

\bibitem[Iben et al. (1967)]{iben67}
Iben, I. Jr, 1967, ARA\&A, 5, 571

\bibitem[]{} Irwin, A.W., 1988, A\&AS, 74, 145
\bibitem[]{} Ishigaki, M.N., Aoki, W., Chiba, M. 2013, ApJ, 771, 67

\bibitem[Johnson et al.(2014)]{johnson14}
Johnson CI, Rich RM, Kobayashi C, Kunder A, Koch A. 2014, AJ, 148, 67
\bibitem[J\"onsson et al. (2017)]{jonsson17}
J\"onsson, H., Ryde, N, Schultheis, M., Zoccali, M. 2017,
A\&A, 600, 2

\bibitem[Karakas \& Lattanzio (2014)]{karakaslattanzio14}
Karakas, A.I., Lattanzio, J.C. 2014, PASA, 31, 30
\bibitem[]{} Kobayashi, C., Umeda, H., Nomoto, K., Tominaga, N., 
Ohkubo, T. 2006, ApJ, 643, 1145


\bibitem[Lamb et al. (2017)]{lamb17}
Lamb, M., Venn, K., Andersen, D. et al. 2017, MNRAS, 465, 3536
\bibitem[]{} Lecureur, A., Hill, V., Zoccali, M., Barbuy, B., et al. 2007,
A\&A, 465, 799

\bibitem[McWilliam(2016)]{mcwilliam16}
McWilliam, A., 2016, PASA,  33, 40



\bibitem[]{} Mel\'endez, J., Asplund, M., Alves-Brito, A. et al.
2008, 484, L21
\bibitem[]{} Mikolaitis, S., Hill, V., Recio-Blanco, A. et al. 2014, A\&A, 572, A33
\bibitem[]{} Mishenina, T.V., Gorbaneva, T.I., Basak, N. Yu. et al. 2011, Astr. Rep. 55, 689
\bibitem[]{} Momany, Y., Vandame, B., Zaggia, S., et al. 2001, A\&A, 379, 436
\bibitem[]{} Ness, M., Freeman, K., Athanassoula, E., et al. 2013, MNRAS, 430, 836
\bibitem[]{} Nissen, P.E., Schuster, W.J. 2011, A\&A, 530, A15
\bibitem[]{}Nomoto, K., Tominaga, N., Umeda, H., Kobayashi, C.,
 Maeda, K. 2006, Nuclear Physics A, 777, 424

\bibitem[]{} Nomoto, K., Kobayashi, C., Tominaga, N. 2013, ARA\&A, 51, 457
\bibitem[]{} Pettini, M., Ellison, S.L., Steidel, C.C., Bowen, D.V. 1999, ApJ, 510, 576


\bibitem[]{} Prochaska, J.S., Naumov, S.O., Carney, B.W., McWilliam, A.,
Wolfe, A.M. 2000, AJ, 120, 2513


\bibitem[Rafelski et al. (2012)]{rafelski12}
Rafelski, M., Wolfe, A.M., Prochaska, J.X. et al. 
2012, ApJ, 755, 89
\bibitem[Rafelski et al. (2014)]{rafelski14}
Rafelski, M., Neeleman, M., Fumagalli, M. et al.
2014, ApJ, 782, L29

\bibitem[]{}
Ram\'{\i}rez, I.,  Mel\'endez, J. 2005, ApJ, 626, 465
\bibitem[]{}
Reddy, B. E., Lambert, D. L., Allende Prieto, C. 2006, MNRAS, 367, 1329
\bibitem[]{} Renzini, A., D'Antona, F., Cassisi, S. et al. 2015,
MNRAS, 454, 4197
\bibitem[]{} Rich, R.M., Origlia, L., Valenti, E. 2012, ApJ, 746, 59
\bibitem[]{} Rojas-Arriagada, A., Recio-Blanco, A., de Laverny, P., et al.
2017, A\&A, 601, 140
\bibitem[]{} Ryde, N., Gustafsson, B., Edvardsson, B. et al. 2010,
A\&A, 509, 20
\bibitem[Saito et al. (2009)]{saito09}
Saito, Y.-J., Takada-Hidai, M., Honda, S., Takeda, Y. 2009, PASJ, 61, 549
\bibitem[Schiavon et al. (2017)]{schiavon17}
Schiavon, R.P., Zamora, O., Carrera, R. et al. 2017, MNRAS, 465, 501
\bibitem[Schlafly \& Finkbeiner(2011)]{Schlafly11}
Schlafly, E.F., Finkbeiner, D.P. 2011, ApJ, 737, 103
\bibitem[Schultheis et al. (2017)]{schultheis17}
Schultheis, M., Rojas-Arriagada, A., Garc\'{\i}a-P\'erez, A.E.
et al. 2017, A\&A, 600, A14

\bibitem[Siqueira-Mello et~al.(2016)]{siqueira-mello16} 
Siqueira-Mello, C., Chiappini, C., Barbuy, B.,  
et~al. 2016, A\&A, 593, A79

\bibitem[Sk\'ulad\'ottir et al.(2017)]{}
Sk\'ulad\'ottir, \'A., Tolstoy, E., Salvadori, S., Hill, V., Pettini, M.
2017, A\&A,  606, A71
\bibitem[Sk\'ulad\'ottir et al.(2018)]{}
Sk\'ulad\'ottir, \'A.,  Salvadori, S., Pettini, M., Tolstoy, E., Hill, V.
2018, A\&A,  in press

\bibitem[Smiljanic et al. (2009)]{smiljanic09}
Smiljanic, R., Gauderon, R., North, P. et al. 2009, A\&A, 502, 267
\bibitem[]{} Sneden, C., Kraft, R.P., Shetrone, M.D. et al. 1997, AJ,  114, 1964
\bibitem[]{} Steffen, M., Prakapavicius, D., Caffau, E.
et al. 2015, A\&A, 583, 57
\bibitem[]{} Tsuji, T. 1973, A\&A, 23, 411
\bibitem[]{} Udalski, A.,  Szymanski M, Kubiak M, et al. 2002, Acta Astron.
52, 217 
\bibitem[]{} Umeda, H., Nomoto, K. 2002, ApJ, 565, 385
\bibitem[]{} Umeda, H., Nomoto, K. 2003, Nature, 422, 871
\bibitem[]{} Umeda, H., Nomoto, K. 2005, ApJ, 619, 427 
\bibitem[]{} Vladilo, G., Abate, C., Yin, J., Cescutti, G., Matteucci, F. 2011,
A\&A, 530, 33










        


        











\bibitem[]{}
Wise, J.H., Turk, M.J., Norman M.L., Abel, T. 2012, ApJ, 745, 50

\bibitem[Woosley \& Weaver]{WW95} 
Woosley, S. E., Weaver, T. A. 1995, ApJS, 101, 181 (WW95)
\bibitem[Woosley et~al.(2002)]{Woosley2002}
Woosley S, Heger A, Weaver TA. 2002. RevModPhys, 74, 1015
\bibitem[]{} Zoccali, M., Lecureur, A., Barbuy, B. et al. 2006, A\&A, 457, L1
\bibitem[]{} Zoccali, M., Hill, V., Lecureur, A., et al. 2008, A\&A, 486, 177
\bibitem[]{} Zoccali, M., Vasquez, S., Gonzalez, OA,
et~al. 2017, A\&A,  599, 12


\end{thebibliography}
\end{document}